\documentclass[%
reprint,
superscriptaddress,
showpacs,
amsmath,amssymb,
aps,
pra,
]{revtex4-1}

\usepackage{graphicx}% Include figure files
\usepackage{dcolumn}% Align table columns on decimal point
\usepackage{bm}% bold math
\usepackage{lipsum}
\usepackage{float}
\usepackage{dsfont}
\usepackage{subfigure}
\usepackage{framed}
\usepackage{xcolor}
\usepackage[colorlinks=true,bookmarks=false]{hyperref}
\hypersetup{linkcolor=blue,citecolor=red,urlcolor=blue}   %\usepackage[mathlines]{lineno}% Enable numbering of text and display math
\begin{document}

\preprint{APS/123-QED}

\title{Lattice Wigner equation}% Force line breaks with \\

\author{S.Sol\'{o}rzano}
\email{sosergio@ethz.ch} 
\affiliation{ ETH
  Z\"urich, Computational Physics for Engineering Materials, Institute
  for Building Materials, Wolfgang-Pauli-Str. 27, HIT, CH-8093 Z\"urich
  (Switzerland)}

\author{M. Mendoza}
%\email{mmendoza@ethz.ch} 
\affiliation{ ETH
  Z\"urich, Computational Physics for Engineering Materials, Institute
  for Building Materials, Wolfgang-Pauli-Str. 27, HIT, CH-8093 Z\"urich
  (Switzerland)}

\author{S. Succi}
%\email{succi@iac.cnr.it} 
\affiliation{Istituto per le Applicazioni del Calcolo C.N.R., Via dei Taurini, 19 00185 Rome, Italy
and Institute for Advanced Computational Science, Harvard University}

\author{H. J. Herrmann}
%\email{hjherrmann@ethz.ch} 
\affiliation{ ETH
  Z\"urich, Computational Physics for Engineering Materials, Institute
  for Building Materials, Wolfgang-Pauli-Str. 27, HIT, CH-8093 Z\"urich
  (Switzerland)}

%\altaffiliation[Also at]{Physics Department, XYZ University.}%Lines break automatically or can be forced with \\
%\author{Author 2}%
% \affiliation{%
%  Department of Physics ETH-Z\"{u}rich
% }%

%\date{April 1, 2015}

\begin{abstract}
We present a numerical scheme to solve the Wigner equation, based on 
a lattice discretization of momentum space.
The moments of the Wigner function are recovered exactly, up to the desired order given by the number 
of discrete momenta retained in the discretisation, which also determines the accuracy of the method. 
The Wigner equation is equipped with an additional collision operator, designed in such a way as to
ensure numerical stability without affecting the evolution of the relevant moments of the Wigner function. 
The lattice Wigner scheme is validated for the case of quantum harmonic and anharmonic 
potentials, showing good agreement with theoretical results. 
It is further applied to the study of the transport properties of one and two
dimensional open quantum systems with potential barriers. 
Finally, the computational viability of the scheme for the 
case of three-dimensional open systems is also illustrated.
\end{abstract}

%\pacs{31.15.-p, 47.11.Qr, 31.15.X- }% PACS, the Physics and Astronomy
\maketitle

\section{Introduction}

The phase-space formulation of quantum mechanics introduced by E.P. Wigner~\cite{PhysRev.40.749} back in 1932,
has known a major surge of interest in the recent years, due to a mounting range of
applications, from quantum chaotic systems~\cite{PhysRevA.62.023612,zurek} and quantum optics~\cite{Alonso:11} 
to ultracold atoms ~\cite{1998PhT51d22L}, for which the Wigner function has been
experimentally reconstructed and measured via tomographic techniques. 

From a theoretical perspective, the Wigner formulation is particularly appealing, because,
by treating position and momentum as two independent quantities, it provides a close bridge 
between quantum mechanics and classical kinetic theory. A bridge which transforms into a complete reconnection in the limit of a vanishing de Broglie length (See \cite{jungel} and references therein).
From a computational viewpoint, however, the Wigner equation is generally demanding and
difficult to handle, its application being often limited to one-dimensional problems. 

Different solution approaches have been proposed in the literature, such as
collocation schemes~\cite{RevModPhys.62.745,doi:10.1137/0727003,doi:10.1137/0732084}, 
semiclassical methods~\cite{PhysRevLett.96.070403,doi:10.1063/1.3425881}, 
Montecarlo approaches~\cite{Sellier2014265,Sellier2014589,Sellier20151,doi:10.1142/S0129156401000897}, finite differences~\cite{Kim19992243,MAINS1994149,Dorda201595}, particle methods~\cite{NICLOT1988313,1464-4266-5-3-381,Filinov2008}, and recently an approach in two dimensions has been proposed~\cite{PhysRevA.92.042122}. 
However, the numerical solution of the Wigner equation still stands as a difficult task, especially in
three spatial dimensions.

Formally, the Wigner equation is similar to the Boltzmann equation, which permits us to borrow 
methods developed in computational kinetic theory to solve quantum mechanical problems.
Of particular interest in this respect, is the Lattice Boltzmann (LB) method, a descendant 
of the lattice gas cellular gas automata\cite{PhysRevLett.56.1505,Wolfram1986}which was originally introduced as an alternative to the discretisation of the Navier-Stokes equations of continuum fluid mechanics \cite{PhysRevLett.61.2332,BENZI1992145,0295-5075-8-6-005}.

Over the years, LB has been adapted to fields as diverse as 
quantum mechanics~\cite{SUCCI1993327, PhysRevLett.113.096402}, relativistic hydrodynamics~\cite{PhysRevLett.105.014502}, 
classical electrodynamics~\cite{PhysRevE.82.056708}, and general relativity \cite{PhysRevE.93.023303}.  
For a recent review, see \cite{EPL2038}. 
The approach is in general computationally efficient and flexible, due to the local 
character of the lattice Boltzmann equation and the fact that information always 
propagates along straight characteristics (light-cones).

In this work, we formulate a lattice Wigner model, borrowing ideas and techniques from lattice 
Boltzmann schemes, namely, the use of a quadrature to reduce the momentum space 
to a small set of representative vectors, thus leading to substantial computational savings.

Even though our work is focussed on the collisionless Wigner equation, the present Lattice Wigner scheme
includes a collision term, for the purpose of numerical stability ~\cite{Furtmaier20161015}. 

Note that  the collision operator is implemented in such a way as to preserve the dynamics of the Wigner function, i.e. the moments correctly reproduced by the numerical quadrature do not experience any dissipation. 

For systems exhibiting genuinely physical dissipation, such constraint can be readily removed, 
so that only the conserved moments are conserved, while the non-conserved ones are indeed affected by dissipative effects. 

The approach is validated for the case of both harmonic and anharmonic quantum oscillators and then applied 
to the transport properties of one and two-dimensional driven open quantum systems. 
Finally, we also show the capability of the model to handle 3D systems with soft potentials.    

This paper is organized a follows: in section \ref{wignerSection}, an introduction to the Wigner 
formalism is given, in section \ref{model} the lattice Wigner model is derived in detail. 
In section \ref{validation}, the model is validated and in section \ref{odqs} it is applied to 
driven open quantum systems. Finally in section \ref{conclusion} the main findings and conclusions are summarized.

\section{Wigner Formulation\label{wignerSection}}

In this section, we provide the basic details about the Wigner formulation, for a comprehensive account see Ref.~\cite{HILLERY1984121}. 
The Wigner formalism is a kinetic formulation of quantum mechanics physically equivalent 
to the Schr\"{o}dinger representation \cite{doi:10.1119/1.1445404}. 
The Wigner formulation, however, is very different, as it treats both position and momenta as independent
variables, like in classical Hamiltionian dynamics and kinetic theory.

The Wigner function is defined as
\begin{align}
W(\mathbf{q},\mathbf{p},t)&=\frac{1}{(2\pi\hbar)^{d}}\mathcal{W}(\hat{\rho})\nonumber\\
&=\frac{1}{(2\pi\hbar)^{d}}\int_{\infty}^{\infty}\mathbf{dy} \rho(\mathbf{q}-\mathbf{y}/2,\mathbf{q}+\mathbf{y}/2) e^{i\mathbf{p}\cdot\mathbf{y}/\hbar},
\label{wignerTransformEq}
\end{align}
where $\rho(\mathbf{x},\mathbf{x'})$ is the real space representation of the density matrix of the 
quantum system under consideration, $d$ is the dimensionality of the system and 
the Weyl transform, $\mathcal{W}(\cdot)$, of a quantum mechanical operator $\hat{O}$ is defined as:
\begin{equation} 
\tilde{O}(\mathbf{q},\mathbf{p})=\mathcal{W}(\hat{O})=\int e^{i\mathbf{p}\cdot\mathbf{y}/\hbar}\langle \mathbf{q}-\mathbf{y}/2|\hat{O}|\mathbf{q}+\mathbf{y}/2\rangle \mathbf{dy}.
\label{weylTransformEq}
\end{equation}
In general, $W(\mathbf{q},\mathbf{p})$ is real and normalised in phase-space, i.e 
$\int \mathbf{dp\; dq} W(\mathbf{p},\mathbf{q},t)=1$. 
However, due to quantum interference effects, it is not positive semidefinite, and consequently, it 
cannot be regarded as a proper distribution function, but rather as a {\it quasi}-distribution. 

Expectation values of a physical observable $\hat{O}$ are obtained through the prescription:
\begin{equation}
\operatorname{tr}({\hat{\rho}\hat{O}})=\int \mathbf{dpdq} \tilde{O}(\mathbf{q},\mathbf{p})W(\mathbf{p},\mathbf{q},t).
\end{equation}
The moments of the Wigner function with respect to the momentum variable are defined as 
\begin{equation} 
\Pi(W)^{n}_{\alpha_{1},...,\alpha_{n}}=\int \mathbf{dp}p_{\alpha_{1}}...p_{\alpha_{n}}W(\mathbf{q},\mathbf{p}),
\label{momentsdef0}
\end{equation}
where $n$ indicates the order of the moment and $p_{\alpha_{i}}$ denotes the $\alpha_{i}$ component of the momentum variable. The first two moments $\Pi(W)^{0}$ and $\Pi(W)^{1}_{\alpha_{i}}$ can be identified with the particle density $\rho(\mathbf{x},\mathbf{x})$ and momentum density respectively, whereas the sum of the diagonal terms of $\Pi(W)^{2}_{\alpha_{i}\alpha_{j}}$ is proportional to the kinetic energy density.  

The time evolution of the Wigner function can be obtained as the Weyl transform 
of the Liouville-von Neumann equation, namely:
\begin{equation}
\frac{\partial\hat{\rho}}{\partial t}=\frac{1}{i\hbar}[\hat{H},\hat{\rho}],\nonumber
\end{equation}
where $\hat{H}=\frac{\hat{\mathbf{p}}^{2}}{2m}+\hat{V}(\mathbf{x})$ is the Hamiltonian of the system. 

The result is known as the Wigner equation and it reads as follows:
\begin{equation}
\frac{\partial W}{\partial t}+\frac{\mathbf{p}}{m}\cdot\nabla W+\Theta[V]W=0,
\label{wignerEq}
\end{equation}  
where $\Theta[V]W$ can be written as
\begin{align}
&\Theta[V]W=\int_{-\infty}^{\infty} \delta[V](\mathbf{q},\mathbf{p}-\mathbf{p}^{\prime})W(\mathbf{q},\mathbf{p}^{\prime})\mathbf{dp}^{\prime}\\
&\delta[V](q,p)=\nonumber\\
&\frac{i}{2\pi\hbar^{2}}\int_{-\infty}^{\infty}(V(\mathbf{q}-\mathbf{y}/2)-V(\mathbf{q}+\mathbf{y}/2))e^{i\mathbf{y}\cdot\mathbf{p}/\hbar}\mathbf{dy}
\label{potentialInEq}
\end{align}
or alternatively 
\begin{equation}
\Theta[V]W=-\sum_{|\mathbf{s}|\in\mathbb{N}_{odd}}\left(\frac{\hbar}{2i}\right)^{|\mathbf{s}|-1}\frac{1}{s !}\frac{\partial^{s} V}{\partial q^{s}}\frac{\partial^{s}W}{\partial p^{s}},
\label{potentialSEq}
\end{equation}
where $\mathbf{s}$ is a vector of non negative integers, $|\mathbf{s}|=\sum_{i=1}^{d}s_{i}$, $\frac{\partial^{s}}{\partial a^{s}}\equiv\Pi_{i=1}^{d}\frac{\partial^{s_{i}}}{\partial{a^{s_{i}}}}$ for $a=\{q,p\}$. 
Finally, it is important to notice that the different terms of the Wigner equation Eq.\eqref{wignerEq} 
can be linked to the different terms of the Liouville-von Neumann equation. 

The convective term arises solely from the kinetic energy term in the Hamiltonian, whereas the 
force term $\Theta[V]W$ originates from the potential energy contribution. 
To be noted that spatial derivatives of the potential at various orders couple to corresponding derivatives 
in momentum space, multiplied by the corresponding power of the Planck's constant $\hbar$. 
Such higher-order terms are responsible for the ``quantumness'' of the Wigner representation 
and the occurrence of negative values due to quantum interference effects. 

% =================================
\section{Lattice Wigner scheme\label{model}}
% =================================

In this section, we introduce the lattice Wigner scheme in two subsequent stages. 
First, the space, time, and velocity discretisation of Eq.\eqref{wignerEq} is described, and 
subsequently, the details on the quadrature in momentum space are presented. 

It is convenient to work in the dimensionless form of Eq.\eqref{wignerEq}. 
Upon the change of variables $q\to l_{0}x$, $p\to m(l_{0}/t_{0})v$, 
$t\to t_{0}\tau$ where $x,v,\tau$ are the new dimensionless 
variables and $l_{0}$, $t_{0}$ are characteristic length and time scales, respectively, 
Eq.\eqref{wignerEq} and Eq.\eqref{potentialSEq} can be written as:
\begin{equation}
\frac{\partial \bar{W}}{\partial \tau} +\mathbf{v}\cdot\nabla_{x}\bar{W}+\Theta{[V]}\bar{W}=0,
\label{wigneDimlessEq}
\end{equation}
and
\begin{equation}
\Theta[V]\bar{W}=-\sum_{|\mathbf{s}|\in\mathbb{N}_{odd}}\left(\frac{H}{2i}\right)^{|\mathbf{s}|-1}\frac{1}{s !}\frac{\partial^{s} \bar{V}}{\partial x^{s}}\frac{\partial^{s}\bar{W}}{\partial v^{s}},
\label{potentialSdimlessEq}
\end{equation}
where $H=\frac{\hbar t_{0}}{ml_{0}^{2}}$, $\bar{V}=\frac{V}{m(l_{0}/t_{0})^2}$ are the 
dimensionless reduced Planck constant and potential terms, respectively. For convenience the relation between physical and dimensionless variables is given in Table.~\ref{VarRelation}
\begin{table}[h]
\begin{ruledtabular}\begin{tabular}{ccc}
Variable & Physical & Lattice\\
\hline
Position & $q$ & $x$\\
Momentum & $p$ & $v$\\
Time & $t$ & $\tau$\\
Reduced Planck constant & $\hbar$ & H\\
Potential & $V$ & $\bar{V}$\\
Wigner function & $W$ & $\bar{W}$
\end{tabular}\end{ruledtabular}
\caption{Relation between physical and lattice variable symbols.}
\label{VarRelation}
\end{table}
\\
\\
The space and time variables of Eq.\eqref{wigneDimlessEq} are discretized simultaneously, that is, first Eq.\eqref{wigneDimlessEq} is formally written as an ordinary differential equation along a $\mathbf{x}+\mathbf{v}\delta\tau\lambda$ line (light-cone), with 
parameter $\lambda\in[0,1]$ and time step $\delta\tau$ 
\begin{equation}
\frac{d \bar{W}}{d\lambda}=-\delta\tau\Theta{[V]}\bar{W}\nonumber,
\label{boltzmanCharacteristicEq}
\end{equation}
This is then integrated leading to: 
\begin{equation}
\bar{W}(\mathbf{x}+\mathbf{v}\delta t,\mathbf{v},t+\delta t)-\bar{W}(\mathbf{x},\mathbf{v},t) = -\delta t\Theta{[V]}\bar{W}.
\label{latticeBoltzmannEq} 
\end{equation}
To be noted that that a first order Taylor series expansion of the l.h.s of Eq.\eqref{latticeBoltzmannEq} 
is consistent with Eq.\eqref{wigneDimlessEq}. 

The velocity space is discretized using quadratures instead of a regular grid approach. 
Besides avoiding the need of a finite cutoff in velocity space, which results in an inaccurate 
computation of the moments of the Wigner distribution, the quadrature approach also 
provides a better discretization of the $\nabla$ operator\cite{Thampi20131}. 

In the present context, discretization by quadrature requires that the moments (Eq.\eqref{momentsdef0}) in the velocity(momentum) space of $\bar{W}$ and $\Theta[V]\bar{W}$ can be calculated exactly. 
This is achieved using a set of $N_{q}$ quadrature vectors and corresponding 
weights $\{\mathbf{v}_{i},w_{i}\}_{i=1}^{N_{q}}$, obeying the consistency relations:
\begin{align}
\Pi(\bar{W})^{n}_{\alpha_{1},\alpha_{1},...,\alpha_{n}}&=\int \mathbf{dv}v_{\alpha_{1}}v_{\alpha_{1}}\cdots v_{\alpha_{n}}\bar{W}(\mathbf{x},\mathbf{v},t)\label{quadratureConstrainEq}\\
&=\sum_{i=0}^{N_{q}}v_{i\alpha_{1}}v_{i\alpha_{1}}\cdots v_{i\alpha_{n}}w_{i}\bar{W}(\mathbf{x},\mathbf{v}_{i},t)\nonumber\\
&=\sum_{i=0}^{N_{q}}v_{i\alpha_{1}}v_{i\alpha_{1}}\cdots v_{i\alpha_{n}}\bar{W}_{i}(\mathbf{x},t),\nonumber
\end{align}
where $\bar{W}_{i}(\mathbf{x},t)=w_{i}\bar{W}(\mathbf{x},\mathbf{v}_{i},t)$ and $v_{i\alpha_{n}}$ denotes the $\alpha_{n}$ component of the i-th velocity vector.
A similar set of equations holds for $\Theta{[V]}\bar{W}$. 

Given a quadrature, Eq.\eqref{latticeBoltzmannEq} is further discretized as 
\begin{equation}
\bar{W}_{i}(\mathbf{x}+\mathbf{v}_{i}\delta t,t+\delta t)-\bar{W}_{i}(\mathbf{x},t) = -\delta t(\Theta{[V]}\bar{W})_{i}.
\label{latticeBoltzmannEq2} 
\end{equation}
Following the lattice Boltzmann nomenclature, the $\bar{W}_{i}$ and $(\Theta{[V]}\bar{W})_{i}$ are termed respectively ``distributions" and ``source distributions". 
Observe that the time evolution of the distributions is given by Eq.\eqref{latticeBoltzmannEq2} and that at every spatial lattice point $\mathbf{x}$, there are $N_{q}$ distributions, from which the moments, such as density $\Pi(\bar{W})^{0}=\rho$ or momentum density $\Pi(\bar{W})^{1}_{\alpha}=\rho\mathbf{u}_{\alpha}$, can be calculated at every time step using Eq.\eqref{quadratureConstrainEq}. 

It is important to notice that, although a discretization by quadrature  
requires no cutoff in velocity space, it does nonetheless involve a ceiling on 
the highest moment for which Eq.\eqref{quadratureConstrainEq} holds.
In other words, it is a truncation in discrete momentum space. 

It is in principle possible to use Eq.\eqref{latticeBoltzmannEq2} to track the time evolution of the moments 
of the Wigner function under the action of a specified potential. However, it was shown in Ref.~\cite{Furtmaier20161015} that the resulting structure of the forcing term leads to numerical instabilities. To address this problem, the lattice Wigner model is introduced as 
\begin{align}
%&\bar{W}_{i}(\mathbf{x}+\mathbf{v}_{i}\delta t,t+\delta t)-\bar{W}_{i}(\mathbf{x},t) = \delta t\Omega_{i}+\delta tS_{i} +\frac{\delta t^{2}}{2}D_{i}S_{i}\nonumber\\ $\frac{\delta t^{2}}{2}D_{i}S_{i}$
&\bar{W}_{i}(\mathbf{x}+\mathbf{v}_{i}\delta t,t+\delta t)-\bar{W}_{i}(\mathbf{x},t) = \delta t\Omega_{i}+\delta tS_{i} +\nonumber\\
&\frac{\delta t}{2}\left(S_{i}(\mathbf{x},t)-S_{i}(\mathbf{x}-\mathbf{v}_{i}\delta t,t-\delta t)\right)\label{lwignerModelEq} \\
&\Omega_{i} = -\frac{1}{\tau_{w}}(\bar{W}_{i}(\mathbf{x},t)-\bar{W}_{i}^{eq}(\mathbf{x},t))\nonumber
\end{align}
where $S_{i}=-(\Theta{[V]}\bar{W})_{i}$ and $\bar{W}^{eq}$ is an artificial ``equilibrium'' distribution such that $\Pi(\bar{W}^{eq})^{n}_{\alpha_{1},\alpha_{1},...,\alpha_{n}}=\Pi(\bar{W})^{n}_{\alpha_{1},\alpha_{1},...,\alpha_{n}}$ for $n\leq N_{\Pi}$. $\tau_{w}>0$ and $N_{\Pi}\in\mathbb{N}$ are model parameters. It is interesting to observe that a formal approach to derive Eq.~\eqref{lwignerModelEq} similar to that presented it Ref.\cite{PhysRevLett.111.090601} for the Lattice Boltzmann equation may be possible. 

Compared to Eq.\eqref{latticeBoltzmannEq2}, Eq.\eqref{lwignerModelEq} exhibits two additional terms. The last one eliminates first-order discretization artifacts~\cite{Shi20081568,PhysRevE.93.043316}, while the first, $\Omega_{i}$, is a regularizing artificial collision term. Since $\Omega$ is a relaxation-type collision term, its use is allowed because it preserves the positive semidefinite character of the density matrix that underlies the Wigner function~\cite{jungel,PhysRevB.88.035401}.  Its role is to improve the stability of the numerical scheme by inducing selective numerical dissipation without directly affecting the dynamics of the first $n\leq N_{\Pi}$ moments of the Wigner equation. This can be seen as follows, let us consider the Taylor expansion up to second order of Eq.\eqref{lwignerModelEq}, namely

\begin{equation} 
%D_{i}\bar{W}_{i}+\frac{\delta t}{2}D_{i}^{2}\bar{W}_{i}=\Omega+S_{i}+\frac{\delta t}{2}D_{i}S_{i},
D_{i}\bar{W}_{i}+\frac{\delta t}{2}D_{i}^{2}\bar{W}_{i}=\Omega+S_{i}+\frac{1}{2}\left(\delta tD_{i}S_{i}-\frac{\delta t^{2}}{2}D_{i}^{2}S_{i}\right),
\label{lwignerTaylorEq}
\end{equation}

By solving for $D_{i}\bar{W}_{i}$ and recursively substituting back in the second term of the l.h.s of Eq.\eqref{lwignerTaylorEq}, it is found that
\begin{align}
% &D_{i}\bar{W}_{i}+\nonumber\\ 
% &\frac{\delta t}{2}D_{i}\left( -\frac{\delta t}{2}D_{i}^{2}\bar{W}_{i} +\Omega+S_{i} +\frac{\delta t}{2}D_{i}S_{i}\right) =\label{lwignerTaylor2Eq}\\
% &\Omega+S_{i} +\frac{\delta t}{2}D_{i}S_{i}.\nonumber
&D_{i}\bar{W}_{i}+\nonumber\\ 
&\frac{\delta t}{2}D_{i}\left( -\frac{\delta t}{2}D_{i}^{2}\bar{W}_{i} +\Omega+S_{i} +\frac{1}{2}\left(\delta tD_{i}S_{i}-\frac{\delta t^{2}}{2}D_{i}^{2}S_{i}\right)\right) =\label{lwignerTaylor2Eq}\\
&\Omega+S_{i} +\frac{1}{2}\left(\delta tD_{i}S_{i}-\frac{\delta t^{2}}{2}D_{i}^{2}S_{i}\right).\nonumber
\end{align}
From Eq.\eqref{lwignerTaylor2Eq}, it can be seen that had the last term of Eq.~\eqref{lwignerModelEq} not been introduced in the definition of the model, there would be an uncompensated source dependent term of order $\delta t$. 
Finally, if the velocity moments of Eq.\eqref{lwignerTaylor2Eq} are calculated, it can be seen that all the contributions 
involving $\Omega_{i}$ vanish, provided that the order of the moment is not larger than $N_{\Pi}$. 
Thus, up to terms of order $O(\delta t^{2})$ and $n\leq N_{\Pi}$,  the resulting set of equations
\begin{align}  
&\frac{\partial}{\partial t}\Pi(\bar{W})^{n}_{\alpha_{1},\alpha_{1},...,\alpha_{n}}+\nabla\cdot\Pi(\bar{W})^{n+1}_{\alpha_{1},\alpha_{1},...,\alpha_{n+1}}\nonumber\\
&=\Pi(\bar{S})^{n}_{\alpha_{1},\alpha_{1},...,\alpha_{n}}+O(\delta t^{2}),
\label{lwignerMomentEq}
\end{align}
is consistent, with the moments of Eq.\eqref{wigneDimlessEq}. 

In summary, Eq.\eqref{lwignerModelEq} approximately solves the Wigner Equation by solving the corresponding 
truncated hierarchy of equations Eq.\eqref{lwignerMomentEq}.

To finalise the model description, a quadrature $\{\mathbf{v}_{i},w_{i}\}_{i=1}^{N_{q}}$ and the explicit expressions for  $\bar{W}_{i}$, $S_{i}$ and $\bar{W}_{i}^{eq}$ are needed.
Since the Wigner function is bounded over the phase space~\cite{doi:10.1119/1.2957889} and only a limited number of moments are required, due to the truncation in Eq.~\eqref{lwignerMomentEq}, an expansion in orthonormal polynomials can be assumed for $\bar{W}$, $S$ and $\bar{W}^{eq}$, from which the expressions of the corresponding distributions can be derived. 

For instance, given a family of polynomials $\{P_{n}(\mathbf{v})\}$, orthonormal under 
the weight function $\omega(\mathbf{v})$, $\bar{W}$ 
can be represented approximately as:
\begin{equation}
\bar{W}(\mathbf{x},\mathbf{v},t)\approx\omega(\mathbf{v})\sum_{n}^{N_{p}}a_{n}(\mathbf{x},t)P_{n}(\mathbf{v}),
\label{wignerPolyRepEq}
\end{equation}    
where $N_{p}$ is the maximum order of the polynomials used in the representation 
and the expansion coefficients are given by
\begin{equation}
a_{n}(\mathbf{x},t)=\int d^{3}\mathbf{v}\bar{W}(\mathbf{x},\mathbf{v},t)P_{n}(\mathbf{v}).
\label{wignerPolyRepCoeffEq}
\end{equation}
It is interesting to note that, since the expansion coefficients are linear combinations of the 
moments of the distribution, this procedure is similar to Grad's method\cite{CPA:CPA3160020403}, 
although not restricted to Hermite polynomials. 

Since any combination of the form $v_{i\alpha_{1}}v_{i\alpha_{1}}\cdots v_{i\alpha_{n}}$ can be represented 
exactly using the set $\{P_{n}(\mathbf{v})\}$, the requirement of Eq.\eqref{quadratureConstrainEq} is 
equivalent to solving the following set of algebraic constraints:
\begin{align}
&\sum_{i=0}^{N_{q}}\omega_{i}P_{n}(\mathbf{v}_{i})P_{m}(\mathbf{v}_{i})=\delta_{n,m}\,\,\,\, \forall n,m\leq N_{p}\label{quadratureEQS}\\
&\mathbf{v}_{i}\in\mathbb{Z}^{d}\,\,\,\, \forall i\nonumber\\
&w_{i}\geq 0\,\,\,\, \forall i,\nonumber
\end{align}
for $\mathbf{v}_{i}$, $\omega_{i}$. 

Technically the constraint $\mathbf{v}_{i}\in\mathbb{Z}^{d}$ is not necessary. 
However, without such constraint, a interpolation would be needed whenever $\mathbf{x}+\mathbf{v}_{i}\delta t$ 
fails to fall on spatial lattice. 
Finally, following a consolidated convention, quadratures in $n$ spatial dimensions 
with $m$ discrete velocity vectors will be designated by $DnQm$. 

Given a solution of Eq.\eqref{quadratureEQS} $\bar{W}_{i}$ can be expressed as
\begin{equation}
\bar{W}_{i}(\mathbf{x},t)=\omega_{i}\sum_{n}^{N_{p}}a_{n}(\mathbf{x},t)P_{n}(\mathbf{v}_{i}),
\label{wignerIthEq}
\end{equation}
and similarly for $S_{i}$ and $\bar{W}^{eq}_{i}$.

In practice, Hermite polynomials are a convenient choice, as they permit the systematic generation
of  lattices in any number of dimensions~\cite{PhysRevLett.97.190601,PhysRevE.79.046701,PhysRevE.73.056702}. 
For example, in one dimension and using Hermite polynomials, $\mathcal{H}_{n}(v;c_{s})$ with weight function $\omega(v;c_{s})=\frac{1}{\sqrt{2\pi c_{s}^{2}}}e^{-\frac{v^{2}}{2c_{s}^{2}}}$ and parameter $c_{s}>0$ \cite{MIZRAHI1975273}, the expressions for $\bar{W}_{i}$, $S_{i}$ and $\bar{W}_{i}^{eq}$ are given by
\begin{align}
&\bar{W}_{i}=\omega_{i}\sum_{n}^{N_{p}}a_{n}(x,t)\mathcal{H}_{n}(v_{i};c_{s})\label{genericithExpansion}\\
&\bar{W}_{i}^{eq}=\omega_{i}\sum_{n}^{N_{\Pi}}a_{n}(x,t)\mathcal{H}_{n}(v_{i};c_{s})\label{genericithExpansionEq}\\
&S_{i}=-\omega_{i}\sum_{n,s}a_{n}(x,t)\sqrt{\frac{(n+s)!}{n!}}\frac{(-H/i)^{s-1}}{c_{s}^{s}s!}\frac{\partial^{s}V}{\partial x^{s}}\mathcal{H}_{n+s}(v_{i};c_{s}).
\label{genericithExpansionS}
\end{align}
Where $H$ is the dimensionless reduced Planck constant. It should be noted that, in general, the condition $N_{\Pi}<N_{p}$ must hold, for otherwise 
$\Omega_{i}$ becomes trivially zero. (For algorithmic details see appendix~\ref{Appendix2}).

Throughout this work Hermite polynomials are in use as they naturally fit the considered problems. However, other choices adapted to particular problems are possible. An example of this, are the generalized polynomials for electronic problems developed in \cite{Rodrigo2016}.

% ===================================  
\section{Validation\label{validation}}
% ===============================

We validate our model, first for the harmonic oscillator and then for the
case of anharmonic potentials with up to sixth order. 

\subsection{Harmonic potential}
 
As a first example to illustrate the lattice Wigner
method described in the previous sections, we consider 
the quantum harmonic oscillator described by the following Hamiltonian Eq.~\eqref{HOEq}:
\begin{equation} 
\hat{H}=\frac{\hat{p}^{2}}{2}+\frac{1}{2}\hat{x}^{2}.
\label{HOEq}
\end{equation}

We track the time propagation of the $\bar{W}_{i}$ distributions from the initial conditions,  
for different choices of the spatial resolution and number of moments $N_{\Pi}$.

The initial condition consists of an equally weighted superposition of the first two eigenstates of the quantum harmonic 
oscillator, $|\phi\rangle=\frac{1}{\sqrt{2}}(|\psi_{0}\rangle+|\psi_{1}\rangle)$. 
Since this state is non stationary, it shows time oscillations all along the evolution. 

The Wigner function corresponding to $|\phi\rangle$ can be calculated from the definition 
Eq.\eqref{wignerTransformEq}, the result being:
\begin{align}
&W_{|\phi\rangle}(x,v)=\frac{e^{-\frac{v^2+x^2}{H}} \left(\sqrt{2} \sqrt{H} x+v^2+x^2\right)}{\pi  H^2}.
\label{homWignerEq}
\end{align}
Observe that if Hermite polynomials are used, $W_{|\phi\rangle}(x,v)$ is already of the form Eq.\eqref{wignerPolyRepEq}. 
It follows that the distributions, $\bar{W}_{i}$, are given by 

\begin{equation}
\bar{W}_{i}=\omega_{i}\frac{e^{-\frac{x^2}{2c_{s}^{2}}}}{\sqrt{2\pi c_{s}^{2}}}
\left(\frac{\mathcal{H}_{2}(v_{i};c_{s})}{2}+\frac{c_{s}^{2}+2c_{s}x+x^{2}}{\sqrt{2}c_{s}^{2}}\mathcal{H}_{1}(v_{i};c_{s})\right)
\end{equation}
where the specific values of $w_{i}$, $v_{i}$ and $c_{s}$ for different lattices 
are given in the Appendix \ref{Appendix1} and $H$ was taken to be numerically equal to $2c_{s}^{2}$.

\begin{figure}[!tb]
\centering
\includegraphics[scale=0.6]{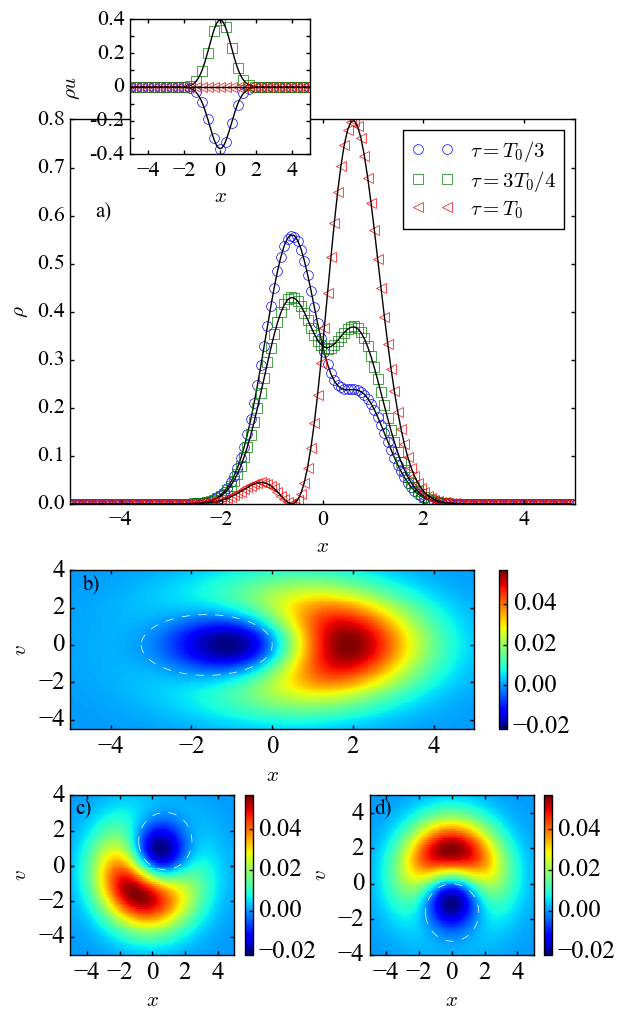}
\caption{(Color on line) a) First moment of the Wigner function (density) for different times being fractions of the oscillation period $T_{0}$. The inset shows the corresponding time evolution for the second moment (velocity density). The symbols denote the simulation and the solid lines the analytical solution. b) Phase space reconstruction of the quantum harmonic oscillator Wigner function at $\tau=T_{0}$. The dashed contour line shows where the Wigner funtion vanishes. c,d) Show the rotation of the harmonic Wigner function for $\tau=T_{0}/3$ and $\tau=T_{0}3/4$,respectively, that is expected from the theory.}
\label{f1}
\end{figure}

\begin{figure}[!t]
\centering
\includegraphics[scale=1]{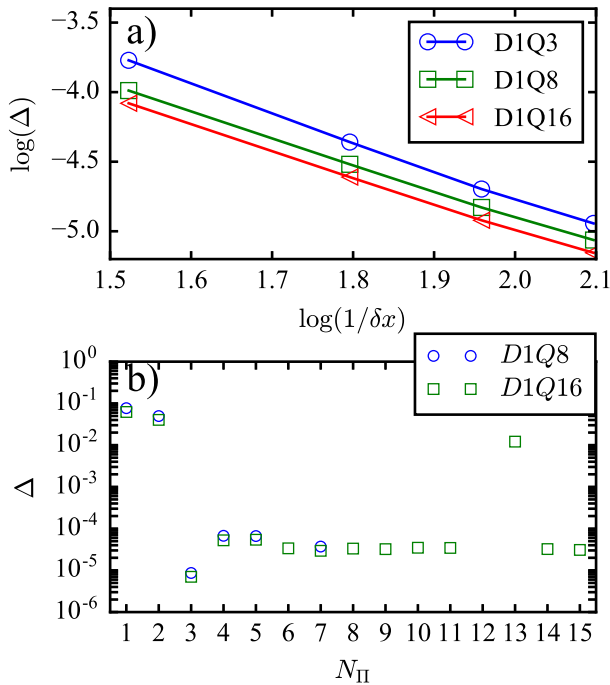}
\caption{(Color online) a) Root mean square error of the density, $\Delta$, for different velocity lattices and spatial resolution using 
an equilibrium function that preserves the first three moments. 
b) Effect on $\Delta$ of using different values of $N_{\Pi}$ for two different lattices with $\delta x=0.008$, $N_{p}=8$ and $N_{p}=16$ for $D1Q8$ and $D1Q16$ respectively.}
\label{f2}
\end{figure}

The results of our simulation using the D1Q3 lattice with a lattice spacing $\delta x=0.06$ 
and a equilibrium function with $N_{\Pi}=3$, are shown in Fig.~\ref{f1}. 
On the upper panel, it can be seen that both the zeroth and first order moments ($\rho$, $\rho u$) 
are correctly propagated and agree with the theoretical values at different times.

From Eq.~\eqref{wignerPolyRepEq}, it is clear that, given the expansion coefficients $a_{n}(\mathbf{x},t)$, it is possible to 
reconstruct an approximation of the Wigner function. These coefficients can be obtained from Eq.\eqref{genericithExpansion}  
as linear combinations of the moments of the Wigner function, which, in turn, can be calculated by means of quadratures. 

The results for the quantum harmonic oscillator are presented on the lower panel of Fig.\ref{f1}, which shows 
the phase-space representation of the Wigner function.
From this figure, a prototypical shape is clearly recognized, including the expected 
nonclassical regions of negative values. 

To quantitatively characterize the present method, we have studied the effects 
of the spatial resolution, lattice configuration and number of preserved moments ($N_{\Pi}$) 

To this end, the root mean square error between the theoretical density and the simulated one 
after a full oscillator period, $T_{0}$, 

\begin{equation}
\Delta=\sqrt{\frac{1}{N_{x}}\sum_{x}(\rho_{theory}(x)-\rho_{sim}(x))^{2}},
\label{errrEq}
\end{equation}

was evaluated for different conditions.
\\
\\\\\
In Fig.\ref{f2} a), the effect of using different lattices and resolution levels is shown; two features 
are apparent, namely that the error $\Delta$ decreases quadratically as a function of 
$1/\delta x$ and that, at a given value of the resolution $\delta x$, schemes with higher 
number of preserved moment provide better results. 

The effect of changing the value of $N_{\Pi}$ is presented in Fig.\ref{f2} b). 
From this figure, an ideal range for $N_{\Pi}$ can be identified. If $N_{\Pi}$ is low, the order of truncation of Eq.\eqref{lwignerMomentEq} leads to a crude approximation which in turn yields large values of $\Delta$. On the other end, if $N_{\Pi}$ is equal to $N_{p}$, the model becomes unstable (this is why we have chosen $N_{\Pi}<N_{p}$ for both lattices), because then the collision term $\Omega_{i}$, Eq.\eqref{lwignerModelEq}, vanishes, which implies no artificial dissipation, hence the onset of stability issues discussed in Ref.\cite{Furtmaier20161015}. The anomalous point $N_{\Pi}=13$, in the D1Q16 case on Fig.\ref{f2} b), may be due to compensated high order modes that reduce the artificial disipation leading to a larger than expected error. That was only observed for the particular case of the harmonic oscillator (and wont be the case for the anharmonic potential).

\begin{figure}[!t]
\centering
\includegraphics[scale=0.54]{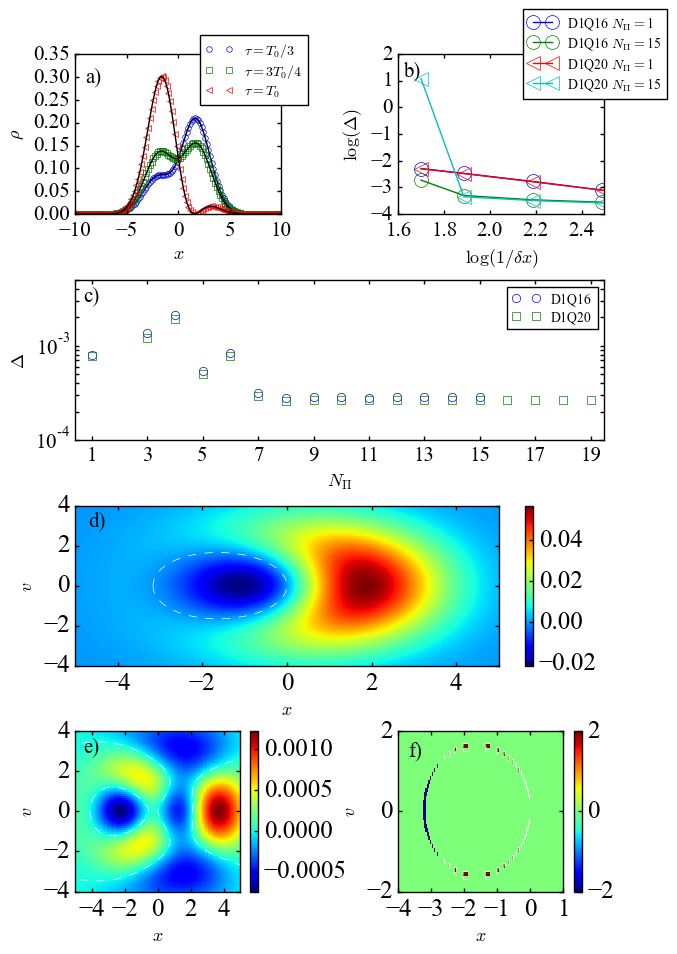}
\caption{(Color online) Results for the anharmonic oscillator with parameters $\alpha=0.1$, $\beta=0.05$. a) Comparison between the density obtained using the Lattice Wigner method and the one obtained directly from the Schr\"{o}dinger equation. (b) Error as a function of resolution, used lattice and number of projections $N_{\Pi}$. It can be seen that as the resolution increases the error saturates and that the error decreases upon increasing the number of projections. (c) Effect on $\Delta$ of using different values of $N_{\Pi}$ for two different lattices with $\delta x=0.008$. (d) Reconstruction of the Wigner function from the anharmonic case, the dashed contour line shows the region where the Wigner function vanishes (e) Difference between the harmonic and anharmonic Wigner functions in phase space. (f) Difference between the signs of the harmonic and anharmonic Wigner functions. A value of $+2$ indicates a region where the harmonic Wigner function is positive and the anharmonic is negative, $-2$ indicates the opposite situation.}
\label{anharmmonicFig}
\end{figure}  

\subsection{Anharmonic potential}
As a second example, we simulate the anharmonic quantum oscillator described by the Hamiltonian 
\begin{equation}
\hat{H}=\frac{\hat{p}^{2}}{2}+\frac{1}{2}\hat{x}^{2}+\alpha \hat{x}^{4}+\beta \hat{x}^{6}\,\,,
\label{anharmonicEq}
\end{equation}
where the parameters $\alpha$ and $\beta$ determine the strength of the anharmonic terms.

As discussed earlier on, anharmonic terms involve genuinely quantum
effects in the forcing expansion described in Eq.~\eqref{potentialSEq}.

Similar to the previous example, the initial condition is taken to be the equal superposition of the first two eigenstates of 
the hamiltonian Eq.\eqref{anharmonicEq} $|\phi\rangle=\frac{1}{\sqrt{2}}\left(|\psi_{0}\rangle+|\psi_{1}\rangle\right)$. 
Here, $|\psi_{0}\rangle$, $|\psi_{1}\rangle$ are obtained by direct diagonalization 
of Eq.~\eqref{anharmonicEq}, using a truncated 
basis set of $50$ eigenvectors, $\varphi_{n}$, from the quantum harmonic oscillator.

The density matrix for this system is given by 
$\hat{\rho}(x,x^{'})=\sum_{m,n}c_{n}c^{*}_{m}\varphi_{m}(x)\varphi_{n}(x')$, where 
the coefficients $c_{n}$ are easily obtained from the diagonalization procedure. 
Given $\hat{\rho}$, the corresponding Wigner function $W_{|\phi\rangle}(x,v)$ was calculated 
with the help of the results in Ref.~\cite{GROENEWOLD1946405}, leading to the following expression:
 
\begin{equation}
W_{|\phi\rangle}(x,v)=\frac{1}{2\pi H}\sum_{n\leq m}\frac{2}{1+\delta_{m,n}}\Re{(c_{n}c_{m}^{*}k_{n,m})},
\label{wignerAnharmoniceq}
\end{equation}
where $\Re(\cdot)$ denotes the real part, the coefficients $k_{m,n}$ are given by
\begin{align}
&k_{m,n}=2(-1)^{\min{(m,n)}}\sqrt{\frac{\min{(m,n)!}}{\max{(m,n)}!}}e^{-\frac{x^{2}+v^{2}}{H}}\nonumber\\
&\left(\frac{2}{H}(x^{2}+v^{2})\right)^{\frac{|m-n|}{2}}L^{|m-n|}_{\min{(m,n)}}\left(\frac{2}{H}(x^{2}+v^{2})\right)\nonumber\\
&e^{(i(m-n)\arctan{(v/x)})}
\label{transitionfucntioneq}
\end{align}
and $L^{m}_{n}$ is the $m$ order $n$ degree associated Laguerre polynomial. 

In order to find the corresponding $\bar{W}_{i}$, the fact is used that each term of Eq.~\eqref{wignerAnharmoniceq} 
can be written as the product of a polynomial and a Gaussian function in the velocity space. 
Once the Gaussian is factored out, the result in Eq.~\eqref{anharmonicWignerExpansioneq} is readily 
cast into the form of Eq.~\eqref{wignerPolyRepEq}, namely 
\begin{equation}  
W(x,v)=\frac{e^{-\frac{v^{2}}{H}}}{\sqrt{2\pi H}}\sum_{n\leq m}\frac{2}{1+\delta_{m,n}}(c_{nm}^{r}\bar{k}_{nm}^{r}-c_{nm}^{i}\bar{k}_{nm}^{i}),
\label{anharmonicWignerExpansioneq}
\end{equation}
In the above,  $c_{nm}=(c_{n}^{r}c_{m}^{r}+c_{n}^{i}c_{m}^{i})+i(c_{n}^{i}c_{m}^{r}-c_{m}^{i}c_{n}^{r})$ 
and $\bar{k}_{mn}=\frac{e^{v^{2}/H}}{2\sqrt{\pi H}}k_{mn}$, where the superscripts $r$ and $i$ denote 
real and imaginary parts, respectively. Since Hermite quadratures are in use, the $\bar{W}_{i}$ follow directly.

% -----------------------------------------------------------

The results for the anharmonic oscillator Eq.\eqref{anharmonicEq} with parameters 
$\alpha=0.1$, $\beta=0.05$ are summarized in Fig.\ref{anharmmonicFig} a), from which it is apparent 
that for mild anharmonicities, the method is able to properly evolve the given initial condition. 
In Fig.\ref{anharmmonicFig} b) the error as a function of the used lattice and resolution 
is reported; the general trend is an error decrease at increasing resolution; it decreases 
as $N_{\Pi}$ increases and $\Delta$ tends to saturate relatively fast. 
In Fig.\ref{anharmmonicFig} c), the behavior of $\Delta$ as function of $N_{\Pi}$ is shown. 
Similarly to the harmonic case, as $N_{\Pi}$ increases, $\Delta$ decreases, until  
it reaches and optimal value ($N_{\Pi}=8$) and then saturates.  

The Wigner function $\bar{W}^{anh}$ was also reconstructed for the anharmonic oscillator Fig.~\ref{anharmmonicFig} d). To observe the quantitative difference between it and the Wigner function of the harmonic oscillator $\bar{W}^{h}$, the difference $\bar{W}^{h}-\bar{W}^{anh}$ is shown in Fig.~\ref{anharmmonicFig} e). Finally it is interesting to notice that for the given levels of anharmonicity the change in the negative region of the Wigner function concentrates on the boundary of the negative region this is shown in Fig.~\ref{anharmmonicFig} f) where the sign difference $\text{sgn}(\bar{W}^{h})-\text{sgn}(\bar{W}^{anh})$ is plotted.

In order to study stronger anharmonic cases, not only larger resolutions, but also more terms in the representation Eq.\eqref{genericithExpansion} of the Wigner function are required because as the strength of the anharmonicity increases, so does the number of terms in Eq.\eqref{anharmonicWignerExpansioneq}. In order to account for them, both the number of polynomials in Eq.\eqref{genericithExpansion} and the size of the velocity lattice needs to be increased.

The effect of the relaxation time of $\tau_{w}$ was also studied. By definition, this parameter controls dissipative effects and consequently, it is not expected to affect the results. However, numerically it was found that this is the case only in the range  $0.56\leq\tau_{w}\leq 5$, which is similar to the allowed range of $\tau_{w}$ in the closely related lattice Boltzmann schemes. This is possibly due to a marginal coupling between high order moments and the ones relevant to the Wigner dynamics.

\subsection{Computational cost}
% ----------------------------------------------------------
\begin{figure}[!t]
\centering
\includegraphics[scale=1]{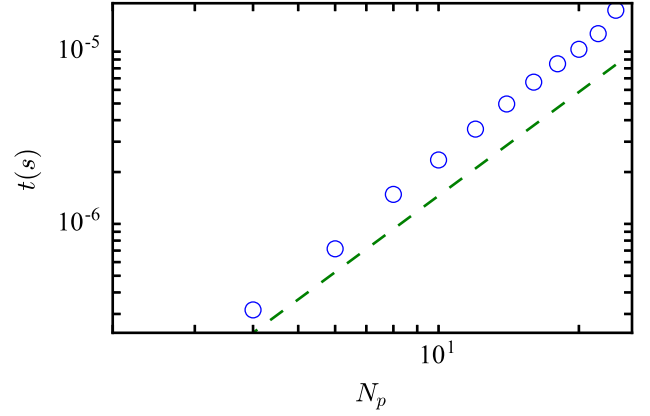}
\caption{(Color online) The symbols show the time it takes to update a single site as a function of the number of polynomials. 
The dashed line shows the scaling $t\sim N_{p}^{2}$. 
The simulations were performed for the quantum harmonic oscillator, in every case  $N_{\Pi}$ was set to the highest value
compatible with numerical stability.}
\label{fcost}
\end{figure}
% --------------------------------------

For an arbitrary problem, it is a priori not known how many polynomials are required to 
give an accurate representation of the Wigner function, Eq.~\eqref{wignerPolyRepEq}. However, similarly to the classical Lattice Boltzmann methods, it is expected that in practice the number of terms in Eq.~\eqref{wignerPolyRepEq} can be minimized if the expected macroscopic velocities $\bar{v}=|\Pi^{1}(\bar{W})/\Pi^{0}(\bar{W})|$ are much smaller than the Lattice speed of sound i.e $|\frac{\bar{v}}{Cs}|\ll1$, and if the expansion Eq.~\eqref{genericithExpansionS} can be truncated on the basis that $|H^{s-1}\partial^{s}_{x}\bar{v}|\ll1$ for certain $s$.   
The number of polynomials, $N_{p}$, determines the smallest lattice that is able to support 
the orthogonality constraints, Eq.~\eqref{quadratureEQS}, and also 
the computational cost of solving the respective problem. 
The scaling of the cost can be estimated by observing that a single update of the complete set 
of lattice points involves four basic steps: 1) the calculation of the expansion coefficients $a_{n}$ in Eq.~\eqref{genericithExpansion}, 2) the update of the source term distributions in Eq.~\eqref{genericithExpansionS}, 3) the update of $\bar{W}^{eq}$ in Eq.~\eqref{wignerIthEq}, and 4) the update of $\bar{W}_{i}$ according to Eq.\eqref{lwignerModelEq}. 

The number of floating point operations ($+,-,\times,/$) required at each step scales 
respectively as $O(N_{p}N_{q})$, $O(N_{s}N_{p}N_{q})$, $O(N_{p}N_{q})$ and $O(1)$, where $N_{s}$ is 
the number of terms in Eq.~\eqref{genericithExpansionS} that are consistent with a cutoff at $s$ in $H$. 

Under a worst-case scenario, i.e. the largest possible $N_{\Pi}$, $N_{\Pi}\sim N_{p}$ and $N_{q}\sim N_{p}$, the 
total cost of updating a single site scales as:
\begin{equation} 
O(N_{s}N_{p}^{2}+N_{p}^{2})
\label{comcost}
\end{equation}
In 1D, $N_{s}$ is effectively $O(1)$ and therefore the cost per site update scales as $O(N_{p}^{2})$. 
This bound was tested and the correpsonding results are reported in Fig.\ref{fcost}, from which it is seen 
that the cost of updating a single site scales like $N_{p}^{2}$. 
The difference with respect to the theoretical value can be accounted for by the time to 
access data, which becomes dominant as the size of the problem is increased.

In 2D, $N_{s}$ scales $O(s^{2})$, and since the number of polynomials and lattice 
vectors also scale quadratically, the update cost per site 
is expected to grow as $O(s^{2}N_{p}^{4})$. For the 3D case $N_{s}$ scales as $O(s^{3})$ and therefore the expected update cost per site is expected to grow as $O(s^{3}N_{p}^{6})$. 

For comparison, the spectral and semispectral methods reported in Ref.~\cite{Furtmaier20161015,doi:10.1137/0727003} scale in $1D$ as $O(N\log{N})$ where $N$ is the number of basis functions. However, this only applies when plane wave basis are used, which are known to introduce numerical artefacts. If an arbitrary basis is used the reported scaling of Ref.~\cite{Furtmaier20161015} becomes $O(N^{2})$. Finally it is interesting to note that Ref.~\cite{PhysRevA.92.042122} reports $O(N\log{N})$ complexity for 2 particles in a single dimension using Fourier methods.

% ----------------------------------------------------------------------
\section{Driven open quantum systems\label{odqs}}
\subsection{1D system}
% ---------------------------------------------------------------
\begin{figure}[t]
\centering
\includegraphics[scale=0.58]{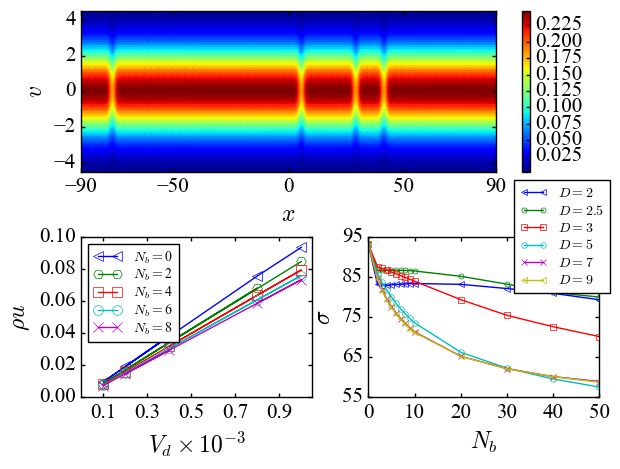}
\caption{(Color online) a): Reconstructed steady state Wigner function. b): 
The second moment of the Wigner function as function of $V_{d}$ in steady state, for different 
number of barriers $N_{b}$. c) $\sigma$ as a function of the number of barriers for different 
inter barrier distances. The system size is set to $400$ (in dimensionless units) and all simulations were performed with a resolution $\delta x=0.004$, with $N_{\Pi}=14$, using a D1Q16 lattice.}
\label{rhouVsVd_rhoVsx}
\end{figure} 

As an application of the proposed model, next we study the dynamics of 
the zeroth and first moment of the Wigner function for a system subject 
to the combined action of an external drive and potential barriers. 
As a model of an homogeneous system, we assume that the initial state is given 
by the following thermal density matrix:
\begin{equation}
\hat{\rho}=\sum|p\rangle\langle p|e^{-\beta p^{2}/2m},
\label{rhoEquilibriumEq}
\end{equation}
where $|p\rangle$ are plane waves, $m$ is the mass of the particle and $\beta$ is the inverse temperature. 

The system is taken to be of finite length $L$, which implies quantization of the allowed momenta. 
However, $L$ is assumed sufficiently large to justify a continuum limit. 

The potential barriers extend throughout the domain according to:
\begin{equation}
V(x)=\frac{v_{0}}{2}\left(\text{erf}\left((x+\delta/2)\xi\right)-\text{erf}\left((x-\delta/2)\xi\right)\right),
\label{barrierVeq}
\end{equation}
where $v_{0}$, $\delta$ and $\xi$ define the height, width and stiffness of the barrier, respectively. 
The barriers were symetrically distributed at the points $\{x_{i}=\pm Di\},\,\,\,i=0,1,\dots,N_{b}$ where $D$ is the interbarrier distance.

The system is driven by the potential $V_{d}(x)=-ax$, where $a$ determines the strength of the forcing, and 
is assumed to be open, i.e. each end of the domain is connected to a fixed 
reservoir, also described by Eq.~\eqref{rhoEquilibriumEq}.

Similar to the previous examples, a lattice Wigner representation of the form Eq.~\eqref{wignerIthEq} is required for the initial condition. In this case, the Wigner transform of Eq.~\eqref{rhoEquilibriumEq} is given by
\begin{equation} 
W(x,v)=\frac{1}{2\pi H}e^{-\frac{v^{2}\bar{\beta}}{2}},
\label{wignerThermalCeq}
\end{equation}
where $\bar{\beta}=\beta m(l_{0}/t_{0})^{2}$. 

Comparing Eq.\eqref{wignerThermalCeq} with the form of the Hermite polynomials weight function, $\omega(v;c_{s})=\frac{1}{\sqrt{2\pi c_{s}^{2}}}e^{-\frac{v^{2}}{2c_{s}^{2}}}$, and using Eq.(\ref{wignerPolyRepEq},\ref{wignerIthEq}) it follows that if $\bar{\beta}$ and $H$ are fixed respectively to $1/c_{s}^{2}$ and $c_{s}$ then only the $a_{0}$ expansion coefficient, that correspond to the constant Hermite polynomial, is requiered. That is, the representation of the initial condition is optimal and the the distributions $\bar{W}_{i}$ are proportional to the weights of the lattice configuration
\begin{equation}
\bar{W}_{i}=\omega_{i}\frac{1}{\sqrt{2\pi}}.
\end{equation}

\begin{figure}[t]
\centering
\includegraphics[scale=1]{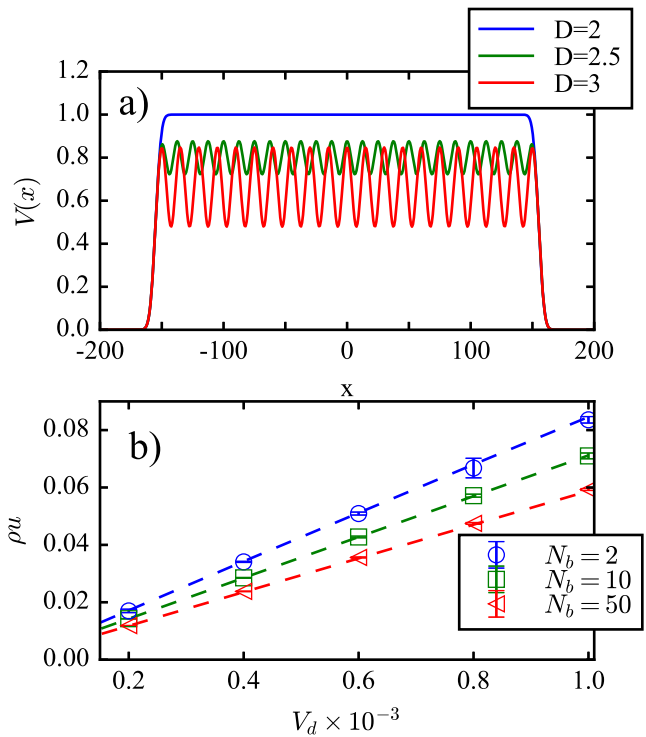}
\caption{(Color online) a) The total potential as a function of the inter barrier separation. For $D=2$ the barriers are close enough such that the total resulting potential acts as a single barrier. As the interbarrier separation increases, the resulting potential exhibits the structure shown for the $D=\{2.5,3\}$ cases b) The symbols show the behavior of $\rho u$, averaged over 50 random samples, as a function of the driving potential. The dashed lines show the behavior of $\rho u$ in the uniformly distributed case with an inter barrier distance $D=8$.}
\label{sigmaVsvars}
\end{figure}

Finally, it is important to observe that the barrier potential Eq.~\eqref{barrierVeq} 
has infinitely many non-zero derivatives, as opposed to the harmonic and anharmonic potentials. 
This implies that a cutoff in Eq.~\eqref{genericithExpansionS} needs to be chosen. 
For the present simulations, the parameters characterising the barriers were fixed as 
$v_{0}=0.4\bar{\beta}^{-1}$, $\delta=2$ and $\xi=1$. 
In this case, the cutoff is taken at $s=9$, since the next contribution, $s=11$, is six orders 
of magnitude smaller than the first order contribution.

The first two moments of the Wigner function were studied for different values of the 
driving force $V_{d}$, number and location of the barriers. 
Fig.~\ref{rhouVsVd_rhoVsx} a) shows the reconstructed steady state Wigner function, $\bar{W}(x,v)$, of 
a system with $V_{d}=10^{-4}$ and 
four barriers randomly located across the domain. 

Similar results were obtained for different configurations of barriers and driving force. 
The first visible feature is that $\bar{W}(x,v)$ shows a number of ``cuts'' along the $v$ axis at given values of $x$. 
These cuts are located at the potential barriers. Along the barriers,
the Wigner function attains lower values as compared to the nearby regions. 
This implies that the density in the cuts is smaller compared to the surroundings. 
 
A second feature is that the Wigner function is nearly translationally invariant in the 
interstitial region between two subsequent cuts, as long as the cuts
are sufficiently far apart, which implies that the density $\rho$ is uniform between cuts. 

Further, from Fig.~\ref{rhouVsVd_rhoVsx} a) it seems that the Wigner distribution 
is symmetric along the $v=0$ axis, although this is not the case. 
The driving potential slightly shifts the distribution, leading to a finite and spatially uniform 
first moment ($\rho u$), which is consistent with the continuity equation
$\frac{\partial\rho}{\partial t}+\nabla\rho u =0$, at steady state. 
Finally, it can be seen that the Wigner function is nowhere negative, first, because the 
reservoir naturally tends to wash out quantum coherence and second, because the ratio between 
the height of the barriers and the thermal energy is about $0.4$, whereas in applications 
such as resonant tunneling diodes, such ratio is about ten~\cite{Dorda201595}. Similarly to the case of strong anharmonicities, to treat systems with higher energy barriers, more terms i.e. polynomials in the representation of the Wigner function (Eq.\eqref{wignerPolyRepEq}) are needed, along with the corresponding increase in the velocity lattice size. For instance, in the case of a resonant tunneling diode, preliminary simulations showed that velocity lattices as large as $D1Q77$ where still not able to recover the negative regions of the system. However, it is expected that by using polynomials adapted to the physical problem one could solve this issue.

From Fig.~\ref{rhouVsVd_rhoVsx} b) it can be seen that the relation between 
the velocity density $\rho u$ and the forcing potential $V_{d}$ is linear 
for a fixed number of barriers, uniformly and symmetrically distributed across the domain. 
Further, Fig.~\ref{rhouVsVd_rhoVsx} b) also implies that, as the number of barriers increases, the 
electric conductivity, $\sigma$, decreases. 

In other terms, the capacity of the system to transport momentum from one end to the other, 
declines with number of barriers. 
To quantify this relation, simulations with a fixed number of barriers, $N_{b}$, but different 
inter-barrier distances, $D$, were performed. 
The results, reported in Fig.\ref{rhouVsVd_rhoVsx} c), show that the overall tendency 
is a decreasing $\sigma$ at increasing $N_b$. 
However, this decrease shows a dependence on the inter-barrier separation $D$. 
For $D=2$ and $D=2.5$, $\sigma$ is nearly constant, whereas for $D \geq 3$ it decreases rapidly with $N_b$. 
Furthermore, $\sigma$ saturates above  $D\geq 5$. 

The above picture can be understood as follows: once the barriers are sufficiently close together, they overlap and the resulting potential is no longer a set of disjoint barriers, but rather a single larger barrier Fig.\ref{sigmaVsvars} a). 

In this case, it is known that all incoming plane waves with energy below the barrier are exponentially attenuated as a function of the barrier length, whereas those with energy above the barrier manage to penetrate, if only with a non-zero reflection probability. 
It follows then that the number of states that can cross the barrier diminishes as the length of the barrier increases thereby limiting the amount of momentum transported across the system, thus leading to an overall decrease of $\sigma$.

\begin{table}[t]
\begin{ruledtabular}\begin{tabular}{lccc}
$D$ & $a$ & $b$ & $c$\\
\hline
5 & $29.1\pm0.6$ & $59.5\pm0.5$ & $0.913\pm0.005$\\
5.5 & $28.7\pm0.7$ & $59.7\pm0.5$ & $0.912\pm0.005$\\
6 & $28.7\pm0.7$ & $59.6\pm0.5$ & $0.913\pm0.005$\\
6.5 & $28.8\pm0.7$ & $59.5\pm0.5$ & $0.914\pm0.005$\\
8 & $28.9\pm0.7$ & $59.4\pm0.5$ & $0.914\pm0.005$
\label{tableFit}
\end{tabular}\end{ruledtabular}
\caption{Individual fitting parameters of Eq.\eqref{sigmaVsNEq} for different interbarrier distances}
\end{table}

When the separation between the barriers is sufficiently large, the system can be approximated 
as a sequence of disjoint barriers. 
If the system was closed, this would imply that, $T$ being the transmission coefficient for a single
incoming plane-wave on a single barrier, the transmission coefficient for $n$ barriers, would 
be $T^{n}$, without any dependence on the inter barrier separation. 
Since this holds for every plane wave contributing to the thermal density matrix, the system as a whole 
is expected to follow a similar trend. 

From the previous picture, it can be inferred that the $\sigma -n$ relation must have a similar 
form for the $D \geq 5$ settings. The semi empirical formula $\sigma = a+bc^{n}$, where $a$, $b$, $c$ are parameters depending on the inter-barrier separation, offers a good fit to the cases $D=5,5.5,6,6.5,8$. 
From Table.\ref{tableFit}, it is apparent that the parameters $a$, $b$ and $c$ are constant within error bars. 
Therefore, for $D \geq 5$, the relation between $\sigma$ and $n$ and $D$, is effectively independent 
of $D$ and given by
\begin{equation}  
\sigma=a+bc^{n}
\label{sigmaVsNEq}
\end{equation}
with $a=28.8\pm 0.1$, $b=59.5\pm 0.1$, $c=0.913\pm 0.001$. 

The intermediate case $2.5<D<5$, when the barriers do not form a single monolithic barrier and the system can no
longer be regarded as a superposition of disjoint subsystems, requires a deeper analysis which is left for future work.

\begin{figure}[t]
\centering
\includegraphics[scale=0.7]{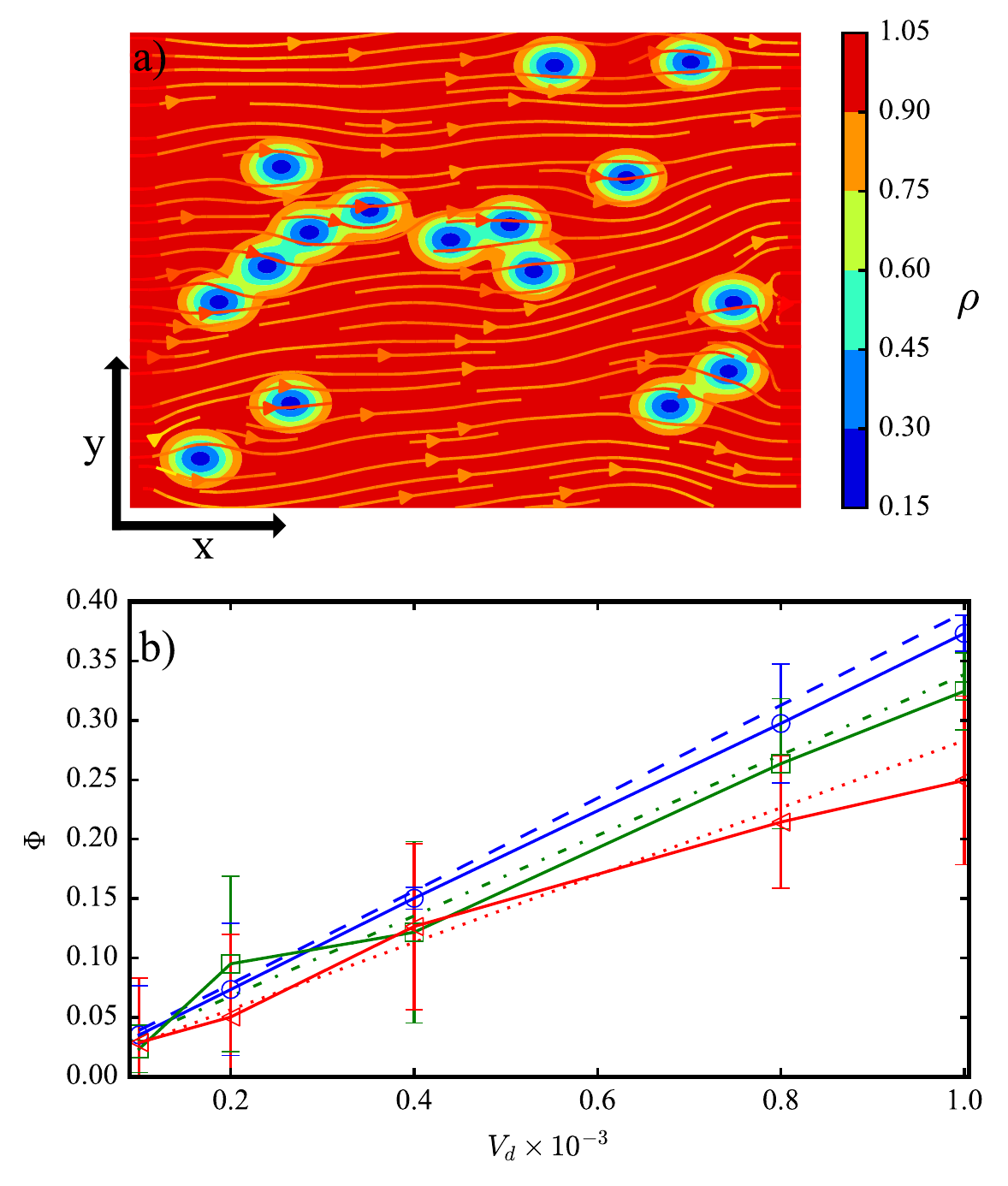}
\caption{(Color online) a) Density map for a 2D system with 16 randomly located barriers. 
The effect of the barriers can be observed on the regions that get depleted (blue color) and on the streamlines that 
bend around them. b) Behavior of $\Phi$ as a function of the driving strength. The red dashed line, blue dot-dashed line and green dotted line correspond to systems where the barriers are arranged in regular grids of $2\times 2$, $3\times 3$ and $4\times 4$  barriers with an inter barrier distance of $D=9$. The circle, square and triangle symbols represent respectively the mean flux of 50 random samples of $2\times 2$, $3\times 3$ and $4\times 4$ randomly located barriers. The solid lines are a guide to the eye showing the trend of $\Phi$ as a function of $V_{d}$ for the case of random barriers.}
\label{2DReRand}
\end{figure}

We have also studied the momentum transport in the presence of a random distribution of barriers. 

Simulations were performed for a fixed number of barriers $N_{b}$, randomly located across the domain. 
The minimum distance between any two barriers was constrained to be larger than 2 lattice sites, in order 
to avoid excessive overlap, leading to an effective single larger barrier instead of two distinct ones. 
The results are presented in Fig.~\ref{sigmaVsvars} b), where for every instance 50 random realisations were considered. 

The main observation is that the relationship between the current $\rho u$ and $V_{d}$ is, on average, the 
same as with uniformly distributed barriers, with an inter-barrier distance $D>5.$ 
This result can be understood as follows; since the barriers are constrained to be far apart, most 
configurations behave as a collection of subsystems. 
This, in turn, implies that $\sigma$ only depends on the number of barriers Eq.\eqref{sigmaVsNEq} 
and, as a consequence, the average relation between $\rho u$ and $V_{d}$ 
does not depart significantly from the case of a regular distribution of barriers.

\subsection{2D system}

The transport properties of a square shaped two-dimensional system of side length $L$, were also studied. 
Open boundary conditions were used at the $x=0$ and $x=L$ ends, while periodic boundary conditions 
are used at the $y=0$ and $y=L$ ends. 
The system is driven by an external potential of the form $V_{d}(\mathbf{x})=-ax$, where 
$a$ controls the strength of the external driving. 
The barriers are described by the potential
\begin{equation}
V(\mathbf{x})=v_{0}e^{-\frac{|\mathbf{x}|^{2}}{2\xi^{2}}},
\end{equation}
where $v_{0}$ determines the height of the barrier and $\xi$ its stiffness. 

The initial state is also given by Eq.~\eqref{rhoEquilibriumEq}, where $|p\rangle$ is assumed to be two dimensional. 
Following calculations similar to the 1D case, the initial condition for the lattice 
Wigner model is given by 
\begin{equation}
\bar{W}_{i}=\omega_{i}\frac{1}{2\pi}.
\end{equation}
The cutoff of Eq.\eqref{potentialSdimlessEq} was set to $s=9$ and the simulations where carried out 
on a $256\times 256$ grid, using the D2Q16 lattice (see Appendix for details).

\begin{figure}[t!]
\centering
\includegraphics[scale=0.3]{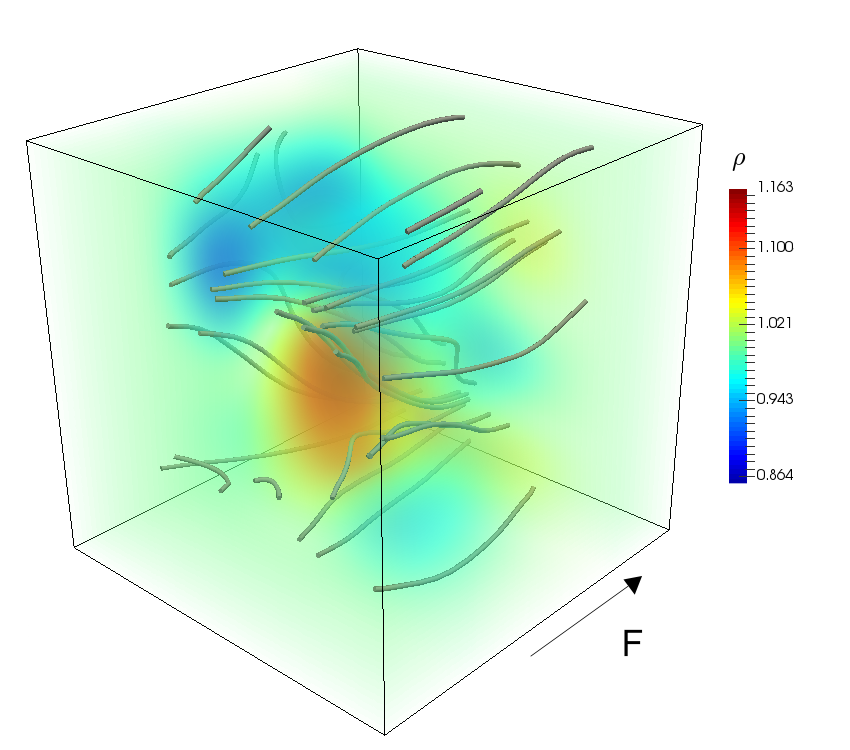}
\caption{(Color online) The figure shows the density $\rho$ and streamlines of $\rho\mathbf{u}$ of an open driven system in 3D. The drive is given by a constant force, $F$, along the $\mathbf{x}$ direction. }
\label{3DWIGNER}
\end{figure}

Similarly to the 1D case, regular and a random settings for the location of the potential barriers were considered. 
Fig.\ref{2DReRand} a) shows a sample result for a simulation with 16 randomly placed barriers. 
The location of the potential barriers can be easily identified through the blue color spots, denoting density depletion. 
Further, it can be seen that the streamlines bend around the potential barriers, similarly to the way 
fluid streamlines turn around obstacles in porous or campylotic media~\cite{campy1,campy2}.

The relation between the flux $\Phi$ (2D analog of $\rho u$ in 1D) and the driving potential 
is presented in  Fig.~\ref{2DReRand} b). 
From this figure, it is seen that the relation $\sigma$ versus $\Phi$ and $V_{d}$ is linear 
when the barriers are regularly organised on a square grid, and that $\sigma$ decreases at increasing number of barriers. Furthermore, when the barriers are randomly placed, the average behavior of $\Phi$ is close to the regular case, as it was also observed in 1D. However as the number of barriers increases, specific realizations can deviate significantly from the regular grid behavior, this can be seen from the error bars of the red triangles in Fig.~\ref{2DReRand} b). 

Finally, for the purpose of showing the viability of the present method also in three spatial dimensions, we 
have simulated a three-dimensional open quantum system. 
The simulation was performed on a $20\times 20\times 20$ lattice, with a D3Q125 velocity set, which was chosen 
because it includes terms of order $H^{2}$ in the force expansion Eq.\eqref{genericithExpansionS}. 
The boundary conditions are open (thermal density matrix) at the planes normal 
to $F$ (See Fig.\ref{3DWIGNER}) and periodic on the remaining boundaries.
In addition to the driving potential generating a force in the $x$ direction, a random potential is included. It is 
modeled as a smooth Gaussian with varying amplitude at different locations in the domain. 
From Fig.\ref{3DWIGNER}, it is seen that the streamlines tend to circumvent the regions of low density, where the 
potential is high, and concentrate in the regions of high density, thus effectively avoiding ``impurities''.
A systematic analysis of the transport properties of this three-dimensional open quantum system is left for future work. 

\section{Conclusion\label{conclusion}}

In this work, a new numerical method to track the time evolution of the Wigner function has been introduced.  
The stability problem previously described in Ref.~\cite{Furtmaier20161015}, is handled through the
inclusion of an artificial collision term, designed in such a way as to preserve the dynamics of 
the relevant moments of the Wigner function.  
The fact of reducing momentum space to a comparatively small set of representative 
momentum vectors, opens up interesting prospects for the 
simulation of one, two and also three dimensional quantum systems.
Preliminary results for 1D systems with regular and random potentials provide 
evidence of linear transport laws which are independent of the barrier configuration for dilute systems. 
In the 2D case, we find the same transport laws at low barrier density, while 
for higher concentrations, deviations from the linear behavior are observed (as shown in Fig.\ref{2DReRand} b) when the barriers are randomly located.
Finally, we also presented a preliminary simulation of a 3D open quantum system, 
to illustrate the ability of the model to handle the three-dimensional Wigner equation.

The computational cost of the method scales polynomially with the number of basis functions.
However, the simulations show that just a few equilibrium moments and comparatively 
small lattices, are often sufficient to obtain reasonably accurate results. 

The present work opens up a number of research directions for the future. 
Technically, the performance can be improved by choosing alternative families of lattice configurations 
and orthonormal polynomials, or by directly designing orthonormal polynomials that fit 
the specific problems under investigation. 
Since our model is computationally viable also in 3D, problems like the heat transport properties 
of three-dimensional semiconductor structures, which are highly relevant to 
the next generation electronics\cite{3Delectronics}, could be studied.
In addition, the method could also be used as a practical tool to explore fundamental issues, such
as the relation between quantum entanglement and the Wigner function in diverse systems\cite{nentanglement,PhysRevA.80.043804} ,or it could be adapted to directly study the time evolution of Hilbert space operators given the relation between these and phase space operators via the Weyl transform.

\section{Acknowledgments}{
We acknowledge financial support from the European Research Council (ERC) Advanced Grant No. 319968-FlowCCS.
}
%kpU96fg
% ===========================================================
\bibliographystyle{apsrev4-1}
\bibliography{bibliography.bib}

\begin{thebibliography}{55}%
\makeatletter
\providecommand \@ifxundefined [1]{%
 \@ifx{#1\undefined}
}%
\providecommand \@ifnum [1]{%
 \ifnum #1\expandafter \@firstoftwo
 \else \expandafter \@secondoftwo
 \fi
}%
\providecommand \@ifx [1]{%
 \ifx #1\expandafter \@firstoftwo
 \else \expandafter \@secondoftwo
 \fi
}%
\providecommand \natexlab [1]{#1}%
\providecommand \enquote  [1]{``#1''}%
\providecommand \bibnamefont  [1]{#1}%
\providecommand \bibfnamefont [1]{#1}%
\providecommand \citenamefont [1]{#1}%
\providecommand \href@noop [0]{\@secondoftwo}%
\providecommand \href [0]{\begingroup \@sanitize@url \@href}%
\providecommand \@href[1]{\@@startlink{#1}\@@href}%
\providecommand \@@href[1]{\endgroup#1\@@endlink}%
\providecommand \@sanitize@url [0]{\catcode `\\12\catcode `\$12\catcode
  `\&12\catcode `\#12\catcode `\^12\catcode `\_12\catcode `\%12\relax}%
\providecommand \@@startlink[1]{}%
\providecommand \@@endlink[0]{}%
\providecommand \url  [0]{\begingroup\@sanitize@url \@url }%
\providecommand \@url [1]{\endgroup\@href {#1}{\urlprefix }}%
\providecommand \urlprefix  [0]{URL }%
\providecommand \Eprint [0]{\href }%
\providecommand \doibase [0]{http://dx.doi.org/}%
\providecommand \selectlanguage [0]{\@gobble}%
\providecommand \bibinfo  [0]{\@secondoftwo}%
\providecommand \bibfield  [0]{\@secondoftwo}%
\providecommand \translation [1]{[#1]}%
\providecommand \BibitemOpen [0]{}%
\providecommand \bibitemStop [0]{}%
\providecommand \bibitemNoStop [0]{.\EOS\space}%
\providecommand \EOS [0]{\spacefactor3000\relax}%
\providecommand \BibitemShut  [1]{\csname bibitem#1\endcsname}%
\let\auto@bib@innerbib\@empty
%</preamble>
\bibitem [{\citenamefont {Wigner}(1932)}]{PhysRev.40.749}%
  \BibitemOpen
  \bibfield  {author} {\bibinfo {author} {\bibfnamefont {E.}~\bibnamefont
  {Wigner}},\ }\href {\doibase 10.1103/PhysRev.40.749} {\bibfield  {journal}
  {\bibinfo  {journal} {Phys. Rev.}\ }\textbf {\bibinfo {volume} {40}},\
  \bibinfo {pages} {749} (\bibinfo {year} {1932})}\BibitemShut {NoStop}%
\bibitem [{\citenamefont {Gardiner}\ \emph {et~al.}(2000)\citenamefont
  {Gardiner}, \citenamefont {Jaksch}, \citenamefont {Dum}, \citenamefont
  {Cirac},\ and\ \citenamefont {Zoller}}]{PhysRevA.62.023612}%
  \BibitemOpen
  \bibfield  {author} {\bibinfo {author} {\bibfnamefont {S.~A.}\ \bibnamefont
  {Gardiner}}, \bibinfo {author} {\bibfnamefont {D.}~\bibnamefont {Jaksch}},
  \bibinfo {author} {\bibfnamefont {R.}~\bibnamefont {Dum}}, \bibinfo {author}
  {\bibfnamefont {J.~I.}\ \bibnamefont {Cirac}}, \ and\ \bibinfo {author}
  {\bibfnamefont {P.}~\bibnamefont {Zoller}},\ }\href {\doibase
  10.1103/PhysRevA.62.023612} {\bibfield  {journal} {\bibinfo  {journal} {Phys.
  Rev. A}\ }\textbf {\bibinfo {volume} {62}},\ \bibinfo {pages} {023612}
  (\bibinfo {year} {2000})}\BibitemShut {NoStop}%
\bibitem [{\citenamefont {Zurek}(2001)}]{zurek}%
  \BibitemOpen
  \bibfield  {author} {\bibinfo {author} {\bibfnamefont {W.~H.}\ \bibnamefont
  {Zurek}},\ }\href@noop {} {\bibfield  {journal} {\bibinfo  {journal}
  {Nature}\ }\textbf {\bibinfo {volume} {412}},\ \bibinfo {pages} {712}
  (\bibinfo {year} {2001})}\BibitemShut {NoStop}%
\bibitem [{\citenamefont {Alonso}(2011)}]{Alonso:11}%
  \BibitemOpen
  \bibfield  {author} {\bibinfo {author} {\bibfnamefont {M.~A.}\ \bibnamefont
  {Alonso}},\ }\href {\doibase 10.1364/AOP.3.000272} {\bibfield  {journal}
  {\bibinfo  {journal} {Adv. Opt. Photon.}\ }\textbf {\bibinfo {volume} {3}},\
  \bibinfo {pages} {272} (\bibinfo {year} {2011})}\BibitemShut {NoStop}%
\bibitem [{\citenamefont {{Leibfried}}\ \emph {et~al.}(1998)\citenamefont
  {{Leibfried}}, \citenamefont {{Pfau}},\ and\ \citenamefont
  {{Monroe}}}]{1998PhT51d22L}%
  \BibitemOpen
  \bibfield  {author} {\bibinfo {author} {\bibfnamefont {D.}~\bibnamefont
  {{Leibfried}}}, \bibinfo {author} {\bibfnamefont {T.}~\bibnamefont {{Pfau}}},
  \ and\ \bibinfo {author} {\bibfnamefont {C.}~\bibnamefont {{Monroe}}},\
  }\href {\doibase 10.1063/1.882256} {\bibfield  {journal} {\bibinfo  {journal}
  {Physics Today}\ }\textbf {\bibinfo {volume} {51}},\ \bibinfo {pages} {22}
  (\bibinfo {year} {1998})}\BibitemShut {NoStop}%
\bibitem [{\citenamefont {{Jungel}}(2009)}]{jungel}%
  \BibitemOpen
  \bibfield  {author} {\bibinfo {author} {\bibfnamefont {A.}~\bibnamefont
  {{Jungel}}},\ }\href@noop {} {\emph {\bibinfo {title} {Transport Equations
  for Semiconductors}}}\ (\bibinfo  {publisher} {Springer},\ \bibinfo {address}
  {Berlin-Heidelberg},\ \bibinfo {year} {2009})\BibitemShut {NoStop}%
\bibitem [{\citenamefont {Frensley}(1990)}]{RevModPhys.62.745}%
  \BibitemOpen
  \bibfield  {author} {\bibinfo {author} {\bibfnamefont {W.~R.}\ \bibnamefont
  {Frensley}},\ }\href {\doibase 10.1103/RevModPhys.62.745} {\bibfield
  {journal} {\bibinfo  {journal} {Rev. Mod. Phys.}\ }\textbf {\bibinfo {volume}
  {62}},\ \bibinfo {pages} {745} (\bibinfo {year} {1990})}\BibitemShut
  {NoStop}%
\bibitem [{\citenamefont {Ringhofer}(1990)}]{doi:10.1137/0727003}%
  \BibitemOpen
  \bibfield  {author} {\bibinfo {author} {\bibfnamefont {C.}~\bibnamefont
  {Ringhofer}},\ }\href {\doibase 10.1137/0727003} {\bibfield  {journal}
  {\bibinfo  {journal} {SIAM Journal on Numerical Analysis}\ }\textbf {\bibinfo
  {volume} {27}},\ \bibinfo {pages} {32} (\bibinfo {year} {1990})},\ \Eprint
  {http://arxiv.org/abs/http://dx.doi.org/10.1137/0727003}
  {http://dx.doi.org/10.1137/0727003} \BibitemShut {NoStop}%
\bibitem [{\citenamefont {Arnold}\ and\ \citenamefont
  {Ringhofer}(1995)}]{doi:10.1137/0732084}%
  \BibitemOpen
  \bibfield  {author} {\bibinfo {author} {\bibfnamefont {A.}~\bibnamefont
  {Arnold}}\ and\ \bibinfo {author} {\bibfnamefont {C.}~\bibnamefont
  {Ringhofer}},\ }\href {\doibase 10.1137/0732084} {\bibfield  {journal}
  {\bibinfo  {journal} {SIAM Journal on Numerical Analysis}\ }\textbf {\bibinfo
  {volume} {32}},\ \bibinfo {pages} {1876} (\bibinfo {year} {1995})},\ \Eprint
  {http://arxiv.org/abs/http://dx.doi.org/10.1137/0732084}
  {http://dx.doi.org/10.1137/0732084} \BibitemShut {NoStop}%
\bibitem [{\citenamefont {Dittrich}\ \emph {et~al.}(2006)\citenamefont
  {Dittrich}, \citenamefont {Viviescas},\ and\ \citenamefont
  {Sandoval}}]{PhysRevLett.96.070403}%
  \BibitemOpen
  \bibfield  {author} {\bibinfo {author} {\bibfnamefont {T.}~\bibnamefont
  {Dittrich}}, \bibinfo {author} {\bibfnamefont {C.}~\bibnamefont {Viviescas}},
  \ and\ \bibinfo {author} {\bibfnamefont {L.}~\bibnamefont {Sandoval}},\
  }\href {\doibase 10.1103/PhysRevLett.96.070403} {\bibfield  {journal}
  {\bibinfo  {journal} {Phys. Rev. Lett.}\ }\textbf {\bibinfo {volume} {96}},\
  \bibinfo {pages} {070403} (\bibinfo {year} {2006})}\BibitemShut {NoStop}%
\bibitem [{\citenamefont {Dittrich}\ \emph {et~al.}(2010)\citenamefont
  {Dittrich}, \citenamefont {Gómez},\ and\ \citenamefont
  {Pachón}}]{doi:10.1063/1.3425881}%
  \BibitemOpen
  \bibfield  {author} {\bibinfo {author} {\bibfnamefont {T.}~\bibnamefont
  {Dittrich}}, \bibinfo {author} {\bibfnamefont {E.~A.}\ \bibnamefont
  {Gómez}}, \ and\ \bibinfo {author} {\bibfnamefont {L.~A.}\ \bibnamefont
  {Pachón}},\ }\href {\doibase 10.1063/1.3425881} {\bibfield  {journal}
  {\bibinfo  {journal} {The Journal of Chemical Physics}\ }\textbf {\bibinfo
  {volume} {132}},\ \bibinfo {pages} {214102} (\bibinfo {year} {2010})},\
  \Eprint {http://arxiv.org/abs/http://dx.doi.org/10.1063/1.3425881}
  {http://dx.doi.org/10.1063/1.3425881} \BibitemShut {NoStop}%
\bibitem [{\citenamefont {Sellier}\ and\ \citenamefont
  {Dimov}(2014{\natexlab{a}})}]{Sellier2014265}%
  \BibitemOpen
  \bibfield  {author} {\bibinfo {author} {\bibfnamefont {J.}~\bibnamefont
  {Sellier}}\ and\ \bibinfo {author} {\bibfnamefont {I.}~\bibnamefont
  {Dimov}},\ }\href {\doibase http://dx.doi.org/10.1016/j.jcp.2014.03.065}
  {\bibfield  {journal} {\bibinfo  {journal} {Journal of Computational
  Physics}\ }\textbf {\bibinfo {volume} {270}},\ \bibinfo {pages} {265 }
  (\bibinfo {year} {2014}{\natexlab{a}})}\BibitemShut {NoStop}%
\bibitem [{\citenamefont {Sellier}\ and\ \citenamefont
  {Dimov}(2014{\natexlab{b}})}]{Sellier2014589}%
  \BibitemOpen
  \bibfield  {author} {\bibinfo {author} {\bibfnamefont {J.}~\bibnamefont
  {Sellier}}\ and\ \bibinfo {author} {\bibfnamefont {I.}~\bibnamefont
  {Dimov}},\ }\href {\doibase https://doi.org/10.1016/j.jcp.2014.05.039}
  {\bibfield  {journal} {\bibinfo  {journal} {Journal of Computational
  Physics}\ }\textbf {\bibinfo {volume} {273}},\ \bibinfo {pages} {589 }
  (\bibinfo {year} {2014}{\natexlab{b}})}\BibitemShut {NoStop}%
\bibitem [{\citenamefont {Sellier}\ \emph {et~al.}(2015)\citenamefont
  {Sellier}, \citenamefont {Nedjalkov},\ and\ \citenamefont
  {Dimov}}]{Sellier20151}%
  \BibitemOpen
  \bibfield  {author} {\bibinfo {author} {\bibfnamefont {J.}~\bibnamefont
  {Sellier}}, \bibinfo {author} {\bibfnamefont {M.}~\bibnamefont {Nedjalkov}},
  \ and\ \bibinfo {author} {\bibfnamefont {I.}~\bibnamefont {Dimov}},\ }\href
  {\doibase https://doi.org/10.1016/j.physrep.2015.03.001} {\bibfield
  {journal} {\bibinfo  {journal} {Physics Reports}\ }\textbf {\bibinfo {volume}
  {577}},\ \bibinfo {pages} {1 } (\bibinfo {year} {2015})},\ \bibinfo {note}
  {an Introduction to Applied Quantum Mechanics in the Wigner Monte Carlo
  Formalism}\BibitemShut {NoStop}%
\bibitem [{\citenamefont {Jacoboni}\ \emph {et~al.}(2001)\citenamefont
  {Jacoboni}, \citenamefont {Brunetti}, \citenamefont {Bordone},\ and\
  \citenamefont {Bertoni}}]{doi:10.1142/S0129156401000897}%
  \BibitemOpen
  \bibfield  {author} {\bibinfo {author} {\bibfnamefont {C.}~\bibnamefont
  {Jacoboni}}, \bibinfo {author} {\bibfnamefont {R.}~\bibnamefont {Brunetti}},
  \bibinfo {author} {\bibfnamefont {P.}~\bibnamefont {Bordone}}, \ and\
  \bibinfo {author} {\bibfnamefont {A.}~\bibnamefont {Bertoni}},\ }\href
  {\doibase 10.1142/S0129156401000897} {\bibfield  {journal} {\bibinfo
  {journal} {International Journal of High Speed Electronics and Systems}\
  }\textbf {\bibinfo {volume} {11}},\ \bibinfo {pages} {387} (\bibinfo {year}
  {2001})}\BibitemShut {NoStop}%
\bibitem [{\citenamefont {Kim}\ and\ \citenamefont {Lee}(1999)}]{Kim19992243}%
  \BibitemOpen
  \bibfield  {author} {\bibinfo {author} {\bibfnamefont {K.-Y.}\ \bibnamefont
  {Kim}}\ and\ \bibinfo {author} {\bibfnamefont {B.}~\bibnamefont {Lee}},\
  }\href {\doibase https://doi.org/10.1016/S0038-1101(99)00168-9} {\bibfield
  {journal} {\bibinfo  {journal} {Solid-State Electronics}\ }\textbf {\bibinfo
  {volume} {43}},\ \bibinfo {pages} {2243 } (\bibinfo {year}
  {1999})}\BibitemShut {NoStop}%
\bibitem [{\citenamefont {Mains}\ and\ \citenamefont
  {Haddad}(1994)}]{MAINS1994149}%
  \BibitemOpen
  \bibfield  {author} {\bibinfo {author} {\bibfnamefont {R.}~\bibnamefont
  {Mains}}\ and\ \bibinfo {author} {\bibfnamefont {G.}~\bibnamefont {Haddad}},\
  }\href {\doibase http://dx.doi.org/10.1006/jcph.1994.1088} {\bibfield
  {journal} {\bibinfo  {journal} {Journal of Computational Physics}\ }\textbf
  {\bibinfo {volume} {112}},\ \bibinfo {pages} {149 } (\bibinfo {year}
  {1994})}\BibitemShut {NoStop}%
\bibitem [{\citenamefont {Dorda}\ and\ \citenamefont
  {Schürrer}(2015)}]{Dorda201595}%
  \BibitemOpen
  \bibfield  {author} {\bibinfo {author} {\bibfnamefont {A.}~\bibnamefont
  {Dorda}}\ and\ \bibinfo {author} {\bibfnamefont {F.}~\bibnamefont
  {Schürrer}},\ }\href {\doibase https://doi.org/10.1016/j.jcp.2014.12.026}
  {\bibfield  {journal} {\bibinfo  {journal} {Journal of Computational
  Physics}\ }\textbf {\bibinfo {volume} {284}},\ \bibinfo {pages} {95 }
  (\bibinfo {year} {2015})}\BibitemShut {NoStop}%
\bibitem [{\citenamefont {Niclot}\ \emph {et~al.}(1988)\citenamefont {Niclot},
  \citenamefont {Degond},\ and\ \citenamefont {Poupaud}}]{NICLOT1988313}%
  \BibitemOpen
  \bibfield  {author} {\bibinfo {author} {\bibfnamefont {B.}~\bibnamefont
  {Niclot}}, \bibinfo {author} {\bibfnamefont {P.}~\bibnamefont {Degond}}, \
  and\ \bibinfo {author} {\bibfnamefont {F.}~\bibnamefont {Poupaud}},\ }\href
  {\doibase http://dx.doi.org/10.1016/0021-9991(88)90053-8} {\bibfield
  {journal} {\bibinfo  {journal} {Journal of Computational Physics}\ }\textbf
  {\bibinfo {volume} {78}},\ \bibinfo {pages} {313 } (\bibinfo {year}
  {1988})}\BibitemShut {NoStop}%
\bibitem [{\citenamefont {Wong}(2003)}]{1464-4266-5-3-381}%
  \BibitemOpen
  \bibfield  {author} {\bibinfo {author} {\bibfnamefont {C.-Y.}\ \bibnamefont
  {Wong}},\ }\href {http://stacks.iop.org/1464-4266/5/i=3/a=381} {\bibfield
  {journal} {\bibinfo  {journal} {Journal of Optics B: Quantum and
  Semiclassical Optics}\ }\textbf {\bibinfo {volume} {5}},\ \bibinfo {pages}
  {S420} (\bibinfo {year} {2003})}\BibitemShut {NoStop}%
\bibitem [{\citenamefont {Filinov}\ \emph {et~al.}(2008)\citenamefont
  {Filinov}, \citenamefont {Bonitz}, \citenamefont {Filinov},\ and\
  \citenamefont {Golubnychiy}}]{Filinov2008}%
  \BibitemOpen
  \bibfield  {author} {\bibinfo {author} {\bibfnamefont {V.~S.}\ \bibnamefont
  {Filinov}}, \bibinfo {author} {\bibfnamefont {M.}~\bibnamefont {Bonitz}},
  \bibinfo {author} {\bibfnamefont {A.}~\bibnamefont {Filinov}}, \ and\
  \bibinfo {author} {\bibfnamefont {V.~O.}\ \bibnamefont {Golubnychiy}},\
  }\enquote {\bibinfo {title} {Wigner function quantum molecular dynamics},}\
  in\ \href {\doibase 10.1007/978-3-540-74686-7_2} {\emph {\bibinfo {booktitle}
  {Computational Many-Particle Physics}}},\ \bibinfo {editor} {edited by\
  \bibinfo {editor} {\bibfnamefont {H.}~\bibnamefont {Fehske}}, \bibinfo
  {editor} {\bibfnamefont {R.}~\bibnamefont {Schneider}}, \ and\ \bibinfo
  {editor} {\bibfnamefont {A.}~\bibnamefont {Wei{\ss}e}}}\ (\bibinfo
  {publisher} {Springer Berlin Heidelberg},\ \bibinfo {address} {Berlin,
  Heidelberg},\ \bibinfo {year} {2008})\ pp.\ \bibinfo {pages}
  {41--60}\BibitemShut {NoStop}%
\bibitem [{\citenamefont {Cabrera}\ \emph {et~al.}(2015)\citenamefont
  {Cabrera}, \citenamefont {Bondar}, \citenamefont {Jacobs},\ and\
  \citenamefont {Rabitz}}]{PhysRevA.92.042122}%
  \BibitemOpen
  \bibfield  {author} {\bibinfo {author} {\bibfnamefont {R.}~\bibnamefont
  {Cabrera}}, \bibinfo {author} {\bibfnamefont {D.~I.}\ \bibnamefont {Bondar}},
  \bibinfo {author} {\bibfnamefont {K.}~\bibnamefont {Jacobs}}, \ and\ \bibinfo
  {author} {\bibfnamefont {H.~A.}\ \bibnamefont {Rabitz}},\ }\href {\doibase
  10.1103/PhysRevA.92.042122} {\bibfield  {journal} {\bibinfo  {journal} {Phys.
  Rev. A}\ }\textbf {\bibinfo {volume} {92}},\ \bibinfo {pages} {042122}
  (\bibinfo {year} {2015})}\BibitemShut {NoStop}%
\bibitem [{\citenamefont {Frisch}\ \emph {et~al.}(1986)\citenamefont {Frisch},
  \citenamefont {Hasslacher},\ and\ \citenamefont
  {Pomeau}}]{PhysRevLett.56.1505}%
  \BibitemOpen
  \bibfield  {author} {\bibinfo {author} {\bibfnamefont {U.}~\bibnamefont
  {Frisch}}, \bibinfo {author} {\bibfnamefont {B.}~\bibnamefont {Hasslacher}},
  \ and\ \bibinfo {author} {\bibfnamefont {Y.}~\bibnamefont {Pomeau}},\ }\href
  {\doibase 10.1103/PhysRevLett.56.1505} {\bibfield  {journal} {\bibinfo
  {journal} {Phys. Rev. Lett.}\ }\textbf {\bibinfo {volume} {56}},\ \bibinfo
  {pages} {1505} (\bibinfo {year} {1986})}\BibitemShut {NoStop}%
\bibitem [{\citenamefont {Wolfram}(1986)}]{Wolfram1986}%
  \BibitemOpen
  \bibfield  {author} {\bibinfo {author} {\bibfnamefont {S.}~\bibnamefont
  {Wolfram}},\ }\href {\doibase 10.1007/BF01021083} {\bibfield  {journal}
  {\bibinfo  {journal} {Journal of Statistical Physics}\ }\textbf {\bibinfo
  {volume} {45}},\ \bibinfo {pages} {471} (\bibinfo {year} {1986})}\BibitemShut
  {NoStop}%
\bibitem [{\citenamefont {McNamara}\ and\ \citenamefont
  {Zanetti}(1988)}]{PhysRevLett.61.2332}%
  \BibitemOpen
  \bibfield  {author} {\bibinfo {author} {\bibfnamefont {G.~R.}\ \bibnamefont
  {McNamara}}\ and\ \bibinfo {author} {\bibfnamefont {G.}~\bibnamefont
  {Zanetti}},\ }\href {\doibase 10.1103/PhysRevLett.61.2332} {\bibfield
  {journal} {\bibinfo  {journal} {Phys. Rev. Lett.}\ }\textbf {\bibinfo
  {volume} {61}},\ \bibinfo {pages} {2332} (\bibinfo {year}
  {1988})}\BibitemShut {NoStop}%
\bibitem [{\citenamefont {Benzi}\ \emph {et~al.}(1992)\citenamefont {Benzi},
  \citenamefont {Succi},\ and\ \citenamefont {Vergassola}}]{BENZI1992145}%
  \BibitemOpen
  \bibfield  {author} {\bibinfo {author} {\bibfnamefont {R.}~\bibnamefont
  {Benzi}}, \bibinfo {author} {\bibfnamefont {S.}~\bibnamefont {Succi}}, \ and\
  \bibinfo {author} {\bibfnamefont {M.}~\bibnamefont {Vergassola}},\ }\href
  {\doibase http://dx.doi.org/10.1016/0370-1573(92)90090-M} {\bibfield
  {journal} {\bibinfo  {journal} {Physics Reports}\ }\textbf {\bibinfo {volume}
  {222}},\ \bibinfo {pages} {145 } (\bibinfo {year} {1992})}\BibitemShut
  {NoStop}%
\bibitem [{\citenamefont {Higuera}\ and\ \citenamefont
  {Succi}(1989)}]{0295-5075-8-6-005}%
  \BibitemOpen
  \bibfield  {author} {\bibinfo {author} {\bibfnamefont {F.~J.}\ \bibnamefont
  {Higuera}}\ and\ \bibinfo {author} {\bibfnamefont {S.}~\bibnamefont
  {Succi}},\ }\href {http://stacks.iop.org/0295-5075/8/i=6/a=005} {\bibfield
  {journal} {\bibinfo  {journal} {EPL (Europhysics Letters)}\ }\textbf
  {\bibinfo {volume} {8}},\ \bibinfo {pages} {517} (\bibinfo {year}
  {1989})}\BibitemShut {NoStop}%
\bibitem [{\citenamefont {Succi}\ and\ \citenamefont
  {Benzi}(1993)}]{SUCCI1993327}%
  \BibitemOpen
  \bibfield  {author} {\bibinfo {author} {\bibfnamefont {S.}~\bibnamefont
  {Succi}}\ and\ \bibinfo {author} {\bibfnamefont {R.}~\bibnamefont {Benzi}},\
  }\href {\doibase http://dx.doi.org/10.1016/0167-2789(93)90096-J} {\bibfield
  {journal} {\bibinfo  {journal} {Physica D: Nonlinear Phenomena}\ }\textbf
  {\bibinfo {volume} {69}},\ \bibinfo {pages} {327 } (\bibinfo {year}
  {1993})}\BibitemShut {NoStop}%
\bibitem [{\citenamefont {Mendoza}\ \emph {et~al.}(2014)\citenamefont
  {Mendoza}, \citenamefont {Succi},\ and\ \citenamefont
  {Herrmann}}]{PhysRevLett.113.096402}%
  \BibitemOpen
  \bibfield  {author} {\bibinfo {author} {\bibfnamefont {M.}~\bibnamefont
  {Mendoza}}, \bibinfo {author} {\bibfnamefont {S.}~\bibnamefont {Succi}}, \
  and\ \bibinfo {author} {\bibfnamefont {H.~J.}\ \bibnamefont {Herrmann}},\
  }\href {\doibase 10.1103/PhysRevLett.113.096402} {\bibfield  {journal}
  {\bibinfo  {journal} {Phys. Rev. Lett.}\ }\textbf {\bibinfo {volume} {113}},\
  \bibinfo {pages} {096402} (\bibinfo {year} {2014})}\BibitemShut {NoStop}%
\bibitem [{\citenamefont {Mendoza}\ \emph {et~al.}(2010)\citenamefont
  {Mendoza}, \citenamefont {Boghosian}, \citenamefont {Herrmann},\ and\
  \citenamefont {Succi}}]{PhysRevLett.105.014502}%
  \BibitemOpen
  \bibfield  {author} {\bibinfo {author} {\bibfnamefont {M.}~\bibnamefont
  {Mendoza}}, \bibinfo {author} {\bibfnamefont {B.~M.}\ \bibnamefont
  {Boghosian}}, \bibinfo {author} {\bibfnamefont {H.~J.}\ \bibnamefont
  {Herrmann}}, \ and\ \bibinfo {author} {\bibfnamefont {S.}~\bibnamefont
  {Succi}},\ }\href {\doibase 10.1103/PhysRevLett.105.014502} {\bibfield
  {journal} {\bibinfo  {journal} {Phys. Rev. Lett.}\ }\textbf {\bibinfo
  {volume} {105}},\ \bibinfo {pages} {014502} (\bibinfo {year}
  {2010})}\BibitemShut {NoStop}%
\bibitem [{\citenamefont {Mendoza}\ and\ \citenamefont
  {Mu\~noz}(2010)}]{PhysRevE.82.056708}%
  \BibitemOpen
  \bibfield  {author} {\bibinfo {author} {\bibfnamefont {M.}~\bibnamefont
  {Mendoza}}\ and\ \bibinfo {author} {\bibfnamefont {J.~D.}\ \bibnamefont
  {Mu\~noz}},\ }\href {\doibase 10.1103/PhysRevE.82.056708} {\bibfield
  {journal} {\bibinfo  {journal} {Phys. Rev. E}\ }\textbf {\bibinfo {volume}
  {82}},\ \bibinfo {pages} {056708} (\bibinfo {year} {2010})}\BibitemShut
  {NoStop}%
\bibitem [{\citenamefont {Ilseven}\ and\ \citenamefont
  {Mendoza}(2016)}]{PhysRevE.93.023303}%
  \BibitemOpen
  \bibfield  {author} {\bibinfo {author} {\bibfnamefont {E.}~\bibnamefont
  {Ilseven}}\ and\ \bibinfo {author} {\bibfnamefont {M.}~\bibnamefont
  {Mendoza}},\ }\href {\doibase 10.1103/PhysRevE.93.023303} {\bibfield
  {journal} {\bibinfo  {journal} {Phys. Rev. E}\ }\textbf {\bibinfo {volume}
  {93}},\ \bibinfo {pages} {023303} (\bibinfo {year} {2016})}\BibitemShut
  {NoStop}%
\bibitem [{\citenamefont {Succi}(2015)}]{EPL2038}%
  \BibitemOpen
  \bibfield  {author} {\bibinfo {author} {\bibfnamefont {S.}~\bibnamefont
  {Succi}},\ }\href {http://stacks.iop.org/0295-5075/109/i=5/a=50001}
  {\bibfield  {journal} {\bibinfo  {journal} {EPL (Europhysics Letters)}\
  }\textbf {\bibinfo {volume} {109}},\ \bibinfo {pages} {50001} (\bibinfo
  {year} {2015})}\BibitemShut {NoStop}%
\bibitem [{\citenamefont {Furtmaier}\ \emph {et~al.}(2016)\citenamefont
  {Furtmaier}, \citenamefont {Succi},\ and\ \citenamefont
  {Mendoza}}]{Furtmaier20161015}%
  \BibitemOpen
  \bibfield  {author} {\bibinfo {author} {\bibfnamefont {O.}~\bibnamefont
  {Furtmaier}}, \bibinfo {author} {\bibfnamefont {S.}~\bibnamefont {Succi}}, \
  and\ \bibinfo {author} {\bibfnamefont {M.}~\bibnamefont {Mendoza}},\ }\href
  {\doibase http://dx.doi.org/10.1016/j.jcp.2015.11.023} {\bibfield  {journal}
  {\bibinfo  {journal} {Journal of Computational Physics}\ }\textbf {\bibinfo
  {volume} {305}},\ \bibinfo {pages} {1015 } (\bibinfo {year}
  {2016})}\BibitemShut {NoStop}%
\bibitem [{\citenamefont {Hillery}\ \emph {et~al.}(1984)\citenamefont
  {Hillery}, \citenamefont {O'Connell}, \citenamefont {Scully},\ and\
  \citenamefont {Wigner}}]{HILLERY1984121}%
  \BibitemOpen
  \bibfield  {author} {\bibinfo {author} {\bibfnamefont {M.}~\bibnamefont
  {Hillery}}, \bibinfo {author} {\bibfnamefont {R.}~\bibnamefont {O'Connell}},
  \bibinfo {author} {\bibfnamefont {M.}~\bibnamefont {Scully}}, \ and\ \bibinfo
  {author} {\bibfnamefont {E.}~\bibnamefont {Wigner}},\ }\href {\doibase
  http://dx.doi.org/10.1016/0370-1573(84)90160-1} {\bibfield  {journal}
  {\bibinfo  {journal} {Physics Reports}\ }\textbf {\bibinfo {volume} {106}},\
  \bibinfo {pages} {121 } (\bibinfo {year} {1984})}\BibitemShut {NoStop}%
\bibitem [{\citenamefont {Styer}\ \emph {et~al.}(2002)\citenamefont {Styer},
  \citenamefont {Balkin}, \citenamefont {Becker}, \citenamefont {Burns},
  \citenamefont {Dudley}, \citenamefont {Forth}, \citenamefont {Gaumer},
  \citenamefont {Kramer}, \citenamefont {Oertel}, \citenamefont {Park},
  \citenamefont {Rinkoski}, \citenamefont {Smith},\ and\ \citenamefont
  {Wotherspoon}}]{doi:10.1119/1.1445404}%
  \BibitemOpen
  \bibfield  {author} {\bibinfo {author} {\bibfnamefont {D.~F.}\ \bibnamefont
  {Styer}}, \bibinfo {author} {\bibfnamefont {M.~S.}\ \bibnamefont {Balkin}},
  \bibinfo {author} {\bibfnamefont {K.~M.}\ \bibnamefont {Becker}}, \bibinfo
  {author} {\bibfnamefont {M.~R.}\ \bibnamefont {Burns}}, \bibinfo {author}
  {\bibfnamefont {C.~E.}\ \bibnamefont {Dudley}}, \bibinfo {author}
  {\bibfnamefont {S.~T.}\ \bibnamefont {Forth}}, \bibinfo {author}
  {\bibfnamefont {J.~S.}\ \bibnamefont {Gaumer}}, \bibinfo {author}
  {\bibfnamefont {M.~A.}\ \bibnamefont {Kramer}}, \bibinfo {author}
  {\bibfnamefont {D.~C.}\ \bibnamefont {Oertel}}, \bibinfo {author}
  {\bibfnamefont {L.~H.}\ \bibnamefont {Park}}, \bibinfo {author}
  {\bibfnamefont {M.~T.}\ \bibnamefont {Rinkoski}}, \bibinfo {author}
  {\bibfnamefont {C.~T.}\ \bibnamefont {Smith}}, \ and\ \bibinfo {author}
  {\bibfnamefont {T.~D.}\ \bibnamefont {Wotherspoon}},\ }\href {\doibase
  10.1119/1.1445404} {\bibfield  {journal} {\bibinfo  {journal} {American
  Journal of Physics}\ }\textbf {\bibinfo {volume} {70}},\ \bibinfo {pages}
  {288} (\bibinfo {year} {2002})},\ \Eprint
  {http://arxiv.org/abs/http://dx.doi.org/10.1119/1.1445404}
  {http://dx.doi.org/10.1119/1.1445404} \BibitemShut {NoStop}%
\bibitem [{\citenamefont {Thampi}\ \emph {et~al.}(2013)\citenamefont {Thampi},
  \citenamefont {Ansumali}, \citenamefont {Adhikari},\ and\ \citenamefont
  {Succi}}]{Thampi20131}%
  \BibitemOpen
  \bibfield  {author} {\bibinfo {author} {\bibfnamefont {S.~P.}\ \bibnamefont
  {Thampi}}, \bibinfo {author} {\bibfnamefont {S.}~\bibnamefont {Ansumali}},
  \bibinfo {author} {\bibfnamefont {R.}~\bibnamefont {Adhikari}}, \ and\
  \bibinfo {author} {\bibfnamefont {S.}~\bibnamefont {Succi}},\ }\href
  {\doibase https://doi.org/10.1016/j.jcp.2012.07.037} {\bibfield  {journal}
  {\bibinfo  {journal} {Journal of Computational Physics}\ }\textbf {\bibinfo
  {volume} {234}},\ \bibinfo {pages} {1 } (\bibinfo {year} {2013})}\BibitemShut
  {NoStop}%
\bibitem [{\citenamefont {B\"osch}\ and\ \citenamefont
  {Karlin}(2013)}]{PhysRevLett.111.090601}%
  \BibitemOpen
  \bibfield  {author} {\bibinfo {author} {\bibfnamefont {F.}~\bibnamefont
  {B\"osch}}\ and\ \bibinfo {author} {\bibfnamefont {I.~V.}\ \bibnamefont
  {Karlin}},\ }\href {\doibase 10.1103/PhysRevLett.111.090601} {\bibfield
  {journal} {\bibinfo  {journal} {Phys. Rev. Lett.}\ }\textbf {\bibinfo
  {volume} {111}},\ \bibinfo {pages} {090601} (\bibinfo {year}
  {2013})}\BibitemShut {NoStop}%
\bibitem [{\citenamefont {Shi}\ \emph {et~al.}(2008)\citenamefont {Shi},
  \citenamefont {Deng}, \citenamefont {Du},\ and\ \citenamefont
  {Chen}}]{Shi20081568}%
  \BibitemOpen
  \bibfield  {author} {\bibinfo {author} {\bibfnamefont {B.}~\bibnamefont
  {Shi}}, \bibinfo {author} {\bibfnamefont {B.}~\bibnamefont {Deng}}, \bibinfo
  {author} {\bibfnamefont {R.}~\bibnamefont {Du}}, \ and\ \bibinfo {author}
  {\bibfnamefont {X.}~\bibnamefont {Chen}},\ }\href {\doibase
  http://doi.org/10.1016/j.camwa.2007.08.016} {\bibfield  {journal} {\bibinfo
  {journal} {Computers \& Mathematics with Applications}\ }\textbf {\bibinfo
  {volume} {55}},\ \bibinfo {pages} {1568 } (\bibinfo {year} {2008})},\
  \bibinfo {note} {mesoscopic Methods in Engineering and Science}\BibitemShut
  {NoStop}%
\bibitem [{\citenamefont {Debus}\ \emph {et~al.}(2016)\citenamefont {Debus},
  \citenamefont {Mendoza}, \citenamefont {Succi},\ and\ \citenamefont
  {Herrmann}}]{PhysRevE.93.043316}%
  \BibitemOpen
  \bibfield  {author} {\bibinfo {author} {\bibfnamefont {J.-D.}\ \bibnamefont
  {Debus}}, \bibinfo {author} {\bibfnamefont {M.}~\bibnamefont {Mendoza}},
  \bibinfo {author} {\bibfnamefont {S.}~\bibnamefont {Succi}}, \ and\ \bibinfo
  {author} {\bibfnamefont {H.~J.}\ \bibnamefont {Herrmann}},\ }\href {\doibase
  10.1103/PhysRevE.93.043316} {\bibfield  {journal} {\bibinfo  {journal} {Phys.
  Rev. E}\ }\textbf {\bibinfo {volume} {93}},\ \bibinfo {pages} {043316}
  (\bibinfo {year} {2016})}\BibitemShut {NoStop}%
\bibitem [{\citenamefont {Rosati}\ \emph {et~al.}(2013)\citenamefont {Rosati},
  \citenamefont {Dolcini}, \citenamefont {Iotti},\ and\ \citenamefont
  {Rossi}}]{PhysRevB.88.035401}%
  \BibitemOpen
  \bibfield  {author} {\bibinfo {author} {\bibfnamefont {R.}~\bibnamefont
  {Rosati}}, \bibinfo {author} {\bibfnamefont {F.}~\bibnamefont {Dolcini}},
  \bibinfo {author} {\bibfnamefont {R.~C.}\ \bibnamefont {Iotti}}, \ and\
  \bibinfo {author} {\bibfnamefont {F.}~\bibnamefont {Rossi}},\ }\href
  {\doibase 10.1103/PhysRevB.88.035401} {\bibfield  {journal} {\bibinfo
  {journal} {Phys. Rev. B}\ }\textbf {\bibinfo {volume} {88}},\ \bibinfo
  {pages} {035401} (\bibinfo {year} {2013})}\BibitemShut {NoStop}%
\bibitem [{\citenamefont {Case}(2008)}]{doi:10.1119/1.2957889}%
  \BibitemOpen
  \bibfield  {author} {\bibinfo {author} {\bibfnamefont {W.~B.}\ \bibnamefont
  {Case}},\ }\href {\doibase 10.1119/1.2957889} {\bibfield  {journal} {\bibinfo
   {journal} {American Journal of Physics}\ }\textbf {\bibinfo {volume} {76}},\
  \bibinfo {pages} {937} (\bibinfo {year} {2008})},\ \Eprint
  {http://arxiv.org/abs/http://dx.doi.org/10.1119/1.2957889}
  {http://dx.doi.org/10.1119/1.2957889} \BibitemShut {NoStop}%
\bibitem [{\citenamefont {Grad}(1949)}]{CPA:CPA3160020403}%
  \BibitemOpen
  \bibfield  {author} {\bibinfo {author} {\bibfnamefont {H.}~\bibnamefont
  {Grad}},\ }\href {\doibase 10.1002/cpa.3160020403} {\bibfield  {journal}
  {\bibinfo  {journal} {Communications on Pure and Applied Mathematics}\
  }\textbf {\bibinfo {volume} {2}},\ \bibinfo {pages} {331} (\bibinfo {year}
  {1949})}\BibitemShut {NoStop}%
\bibitem [{\citenamefont {Chikatamarla}\ and\ \citenamefont
  {Karlin}(2006)}]{PhysRevLett.97.190601}%
  \BibitemOpen
  \bibfield  {author} {\bibinfo {author} {\bibfnamefont {S.~S.}\ \bibnamefont
  {Chikatamarla}}\ and\ \bibinfo {author} {\bibfnamefont {I.~V.}\ \bibnamefont
  {Karlin}},\ }\href {\doibase 10.1103/PhysRevLett.97.190601} {\bibfield
  {journal} {\bibinfo  {journal} {Phys. Rev. Lett.}\ }\textbf {\bibinfo
  {volume} {97}},\ \bibinfo {pages} {190601} (\bibinfo {year}
  {2006})}\BibitemShut {NoStop}%
\bibitem [{\citenamefont {Chikatamarla}\ and\ \citenamefont
  {Karlin}(2009)}]{PhysRevE.79.046701}%
  \BibitemOpen
  \bibfield  {author} {\bibinfo {author} {\bibfnamefont {S.~S.}\ \bibnamefont
  {Chikatamarla}}\ and\ \bibinfo {author} {\bibfnamefont {I.~V.}\ \bibnamefont
  {Karlin}},\ }\href {\doibase 10.1103/PhysRevE.79.046701} {\bibfield
  {journal} {\bibinfo  {journal} {Phys. Rev. E}\ }\textbf {\bibinfo {volume}
  {79}},\ \bibinfo {pages} {046701} (\bibinfo {year} {2009})}\BibitemShut
  {NoStop}%
\bibitem [{\citenamefont {Philippi}\ \emph {et~al.}(2006)\citenamefont
  {Philippi}, \citenamefont {Hegele}, \citenamefont {dos Santos},\ and\
  \citenamefont {Surmas}}]{PhysRevE.73.056702}%
  \BibitemOpen
  \bibfield  {author} {\bibinfo {author} {\bibfnamefont {P.~C.}\ \bibnamefont
  {Philippi}}, \bibinfo {author} {\bibfnamefont {L.~A.}\ \bibnamefont
  {Hegele}}, \bibinfo {author} {\bibfnamefont {L.~O.~E.}\ \bibnamefont {dos
  Santos}}, \ and\ \bibinfo {author} {\bibfnamefont {R.}~\bibnamefont
  {Surmas}},\ }\href {\doibase 10.1103/PhysRevE.73.056702} {\bibfield
  {journal} {\bibinfo  {journal} {Phys. Rev. E}\ }\textbf {\bibinfo {volume}
  {73}},\ \bibinfo {pages} {056702} (\bibinfo {year} {2006})}\BibitemShut
  {NoStop}%
\bibitem [{\citenamefont {Mizrahi}(1975)}]{MIZRAHI1975273}%
  \BibitemOpen
  \bibfield  {author} {\bibinfo {author} {\bibfnamefont {M.~M.}\ \bibnamefont
  {Mizrahi}},\ }\href {\doibase http://dx.doi.org/10.1016/0771-050X(75)90019-4}
  {\bibfield  {journal} {\bibinfo  {journal} {Journal of Computational and
  Applied Mathematics}\ }\textbf {\bibinfo {volume} {1}},\ \bibinfo {pages}
  {273 } (\bibinfo {year} {1975})}\BibitemShut {NoStop}%
\bibitem [{\citenamefont {{Coelho, Rodrigo C. V.}}\ \emph
  {et~al.}(2016)\citenamefont {{Coelho, Rodrigo C. V.}}, \citenamefont {{Ilha,
  Anderson S.}},\ and\ \citenamefont {{Doria, Mauro M.}}}]{Rodrigo2016}%
  \BibitemOpen
  \bibfield  {author} {\bibinfo {author} {\bibnamefont {{Coelho, Rodrigo C.
  V.}}}, \bibinfo {author} {\bibnamefont {{Ilha, Anderson S.}}}, \ and\
  \bibinfo {author} {\bibnamefont {{Doria, Mauro M.}}},\ }\href {\doibase
  10.1209/0295-5075/116/20001} {\bibfield  {journal} {\bibinfo  {journal}
  {EPL}\ }\textbf {\bibinfo {volume} {116}},\ \bibinfo {pages} {20001}
  (\bibinfo {year} {2016})}\BibitemShut {NoStop}%
\bibitem [{\citenamefont {Groenewold}(1946)}]{GROENEWOLD1946405}%
  \BibitemOpen
  \bibfield  {author} {\bibinfo {author} {\bibfnamefont {H.}~\bibnamefont
  {Groenewold}},\ }\href {\doibase
  http://dx.doi.org/10.1016/S0031-8914(46)80059-4} {\bibfield  {journal}
  {\bibinfo  {journal} {Physica}\ }\textbf {\bibinfo {volume} {12}},\ \bibinfo
  {pages} {405 } (\bibinfo {year} {1946})}\BibitemShut {NoStop}%
\bibitem [{\citenamefont {Mendoza}\ \emph {et~al.}(2013)\citenamefont
  {Mendoza}, \citenamefont {Succi},\ and\ \citenamefont {Herrmann}}]{campy1}%
  \BibitemOpen
  \bibfield  {author} {\bibinfo {author} {\bibfnamefont {M.}~\bibnamefont
  {Mendoza}}, \bibinfo {author} {\bibfnamefont {S.}~\bibnamefont {Succi}}, \
  and\ \bibinfo {author} {\bibfnamefont {H.~J.}\ \bibnamefont {Herrmann}},\
  }\href@noop {} {\bibfield  {journal} {\bibinfo  {journal} {Scientific
  Reports}\ }\textbf {\bibinfo {volume} {3}},\ \bibinfo {pages} {3106 EP }
  (\bibinfo {year} {2013})}\BibitemShut {NoStop}%
\bibitem [{\citenamefont {Debus}\ \emph {et~al.}(2017)\citenamefont {Debus},
  \citenamefont {Mendoza}, \citenamefont {Succi},\ and\ \citenamefont
  {Herrmann}}]{campy2}%
  \BibitemOpen
  \bibfield  {author} {\bibinfo {author} {\bibfnamefont {J.~D.}\ \bibnamefont
  {Debus}}, \bibinfo {author} {\bibfnamefont {M.}~\bibnamefont {Mendoza}},
  \bibinfo {author} {\bibfnamefont {S.}~\bibnamefont {Succi}}, \ and\ \bibinfo
  {author} {\bibfnamefont {H.~J.}\ \bibnamefont {Herrmann}},\ }\href@noop {}
  {\bibfield  {journal} {\bibinfo  {journal} {Scientific Reports}\ }\textbf
  {\bibinfo {volume} {7}},\ \bibinfo {pages} {42350 EP } (\bibinfo {year}
  {2017})}\BibitemShut {NoStop}%
\bibitem [{\citenamefont {Lu}\ and\ \citenamefont
  {Lieber}(2007)}]{3Delectronics}%
  \BibitemOpen
  \bibfield  {author} {\bibinfo {author} {\bibfnamefont {W.}~\bibnamefont
  {Lu}}\ and\ \bibinfo {author} {\bibfnamefont {C.~M.}\ \bibnamefont
  {Lieber}},\ }\href {http://dx.doi.org/10.1038/nmat2028} {\bibfield  {journal}
  {\bibinfo  {journal} {Nat Mater}\ }\textbf {\bibinfo {volume} {6}},\ \bibinfo
  {pages} {841} (\bibinfo {year} {2007})}\BibitemShut {NoStop}%
\bibitem [{\citenamefont {McConnell}\ \emph {et~al.}(2015)\citenamefont
  {McConnell}, \citenamefont {Zhang}, \citenamefont {Hu}, \citenamefont {Cuk},\
  and\ \citenamefont {Vuletic}}]{nentanglement}%
  \BibitemOpen
  \bibfield  {author} {\bibinfo {author} {\bibfnamefont {R.}~\bibnamefont
  {McConnell}}, \bibinfo {author} {\bibfnamefont {H.}~\bibnamefont {Zhang}},
  \bibinfo {author} {\bibfnamefont {J.}~\bibnamefont {Hu}}, \bibinfo {author}
  {\bibfnamefont {S.}~\bibnamefont {Cuk}}, \ and\ \bibinfo {author}
  {\bibfnamefont {V.}~\bibnamefont {Vuletic}},\ }\href
  {http://dx.doi.org/10.1038/nature14293} {\bibfield  {journal} {\bibinfo
  {journal} {Nature}\ }\textbf {\bibinfo {volume} {519}},\ \bibinfo {pages}
  {439} (\bibinfo {year} {2015})}\BibitemShut {NoStop}%
\bibitem [{\citenamefont {Gonzalez}\ \emph {et~al.}(2009)\citenamefont
  {Gonzalez}, \citenamefont {Molina-Terriza},\ and\ \citenamefont
  {Torres}}]{PhysRevA.80.043804}%
  \BibitemOpen
  \bibfield  {author} {\bibinfo {author} {\bibfnamefont {N.}~\bibnamefont
  {Gonzalez}}, \bibinfo {author} {\bibfnamefont {G.}~\bibnamefont
  {Molina-Terriza}}, \ and\ \bibinfo {author} {\bibfnamefont {J.~P.}\
  \bibnamefont {Torres}},\ }\href {\doibase 10.1103/PhysRevA.80.043804}
  {\bibfield  {journal} {\bibinfo  {journal} {Phys. Rev. A}\ }\textbf {\bibinfo
  {volume} {80}},\ \bibinfo {pages} {043804} (\bibinfo {year}
  {2009})}\BibitemShut {NoStop}%
\bibitem [{\citenamefont {Succi}(2001)}]{succi2001lattice}%
  \BibitemOpen
  \bibfield  {author} {\bibinfo {author} {\bibfnamefont {S.}~\bibnamefont
  {Succi}},\ }\href@noop {} {\emph {\bibinfo {title} {The lattice Boltzmann
  equation: for fluid dynamics and beyond}}}\ (\bibinfo  {publisher} {Oxford
  university press},\ \bibinfo {year} {2001})\BibitemShut {NoStop}%
\end{thebibliography}%

\clearpage
\appendix
\section{Lattice specification\label{Appendix1}}
\begin{table}[h]
\caption{D1Q4 Lattice with $c_{s}=0.60625445810016454$}
\begin{ruledtabular}\begin{tabular}{cl}
$\mathbf{v}_{i}$ & $w_{i} $\\
\hline
0 & 0.63664690312607816284434609283846\\
-1,1 & 0.18141458774368577505004149208377\\
-3,3& 0.00026196069327514352778546149699\\
\end{tabular}\end{ruledtabular}
\end{table}

\begin{table}[h]
\caption{D1Q8 Lattice with $c_{s}=1.0658132602705641$}
\begin{ruledtabular}\begin{tabular}{cl}
$\mathbf{v}_{i}$ & $w_{i} $\\
\hline
0 & 0.37428019874212190129215011724318\\
-1,1 & 0.24105344284458452784844296921093\\
-2,2& 0.06434304152476086575379872184362\\
-3,3& 0.00713156628791277339406557854605\\
-4,4&0.00032523057375714836476726255033\\
-5,5&6.6163470389851878681133911638949$\times 10^{-6}$\\
-7,7&3.0508847488049822958363118638543$\times 10^{-9}$\\
\end{tabular}\end{ruledtabular}
\end{table}

\begin{table}[h]
\caption{D1Q10 Lattice with $c_{s}=1.229594448425497$}
\begin{ruledtabular}\begin{tabular}{cl}
$\mathbf{v}_{i}$ & $w_{i} $\\
\hline
 0   & 0.32444899174631946866086595194671\\  
-1,1 & 0.23309081165504033632566413700874\\  
-2,2 & 0.08642582836940192624063184539752\\  
-3,3 & 0.01653989847863324979993319254793\\  
-4,4 & 0.00163342485156222352004541584861\\  
-5,5 & 0.00008333063878279730921268566542\\  
-6,6 & 2.1783167706100240902344965225688$\times 10^{-6}$\\ 
-7,7 & 3.1805869765623071575130276965860$\times 10^{-8}$\\  
-9,9 & 1.0779356826917937931616055896767$\times 10^{-11}$\\  
\end{tabular}\end{ruledtabular}
\end{table} 

\begin{table}[h]
\caption{D1Q16 Lattice with $c_{s}=1.6215048099592275$}
\begin{ruledtabular}\begin{tabular}{cl}
$\mathbf{v}_{i}$ & $w_{i} $\\
\hline
0 & 0.24603212869787232483785340883852\\
-1,1 & 0.20342468717937742901117526034797\\
-2,2& 0.11498446042457243913866706495342\\
-3,3& 0.04443225067964028999644337006636\\
-4,4&0.01173764938741580915572505702913\\
-5,5&0.00211976456798849884644315007219\\
-6,6&0.00026170845228301249011385925086\\
-7,7&0.00002208877826469659955769726449\\
-8,8&1.2745253026359480126112714680367$\times 10^{-6}$\\
-9,9&5.0275261810959383411581297192576$\times 10^{-8}$\\
-10,10&1.3556297819769484757032262002820$\times 10^{-9}$\\
-11,11&2.5012031341031852003279373539252$\times 10^{-11}$\\
-12,12&3.1243604817078012360750317883072$\times 10^{-13}$\\
-13,13&2.9655118189640940365948400709026$\times 10^{-15}$\\
-15,15&5.5758174181938354491800200913102$\times 10^{-19}$\\
\end{tabular}\end{ruledtabular}
\end{table}
\begin{table}[h]
\caption{D1Q20 Lattice with $c_{s}=1.8357424381402594$}
\begin{ruledtabular}\begin{tabular}{cl}
$\mathbf{v}_{i}$ & $w_{i} $\\
\hline
0& 0.21731931022112109059537537887018\\
-1,1& 0.18735357499686018912399983787195\\
-2,2& 0.12004746243830897823022161375249\\
-3,3& 0.05717041140835294313179190148076\\
-4,4& 0.02023564183037203154174508370450\\
-5,5& 0.00532341082536521716813053993040\\
-6,6& 0.00104085519989277817787032717819\\
-7,7& 0.00015125787069729717011289372472\\
-8,8& 0.00001633702528266419558030012576\\
-9,9& 1.3114608069909806258412013955825$\times 10^{-6}$\\
-10,10& 7.8246508661616857867473666191193$\times 10^{-8}$\\
-11,11& 3.4697808952346470123102609636720$\times 10^{-9}$\\
-12,12& 1.1435779630395075964965610648417$\times 10^{-10}$\\
-13,13& 2.8013182217362421082623834683491$\times 10^{-12}$\\
-14,14& 5.0995884226301388644438757982605$\times 10^{-14}$\\
-15,15& 6.9079520892785667788901676695952$\times 10^{-16}$\\
-16,16& 6.8680470174442627832690379600090$\times 10^{-18}$\\
-17,17& 5.7551467186859264824045886746476$\times 10^{-20}$\\
-19,19& 8.3761764243303081227469285304101$\times 10^{-24}$\\
\end{tabular}\end{ruledtabular}
\end{table}

2D lattices and in general $n$ dimensional lattices can be constructed by taking $n$ times the tensor product of the set of vectors and weights of a fixed 1D Lattice. 
For example, the D2Q4 lattice is given by Table.\ref{tableD2Q4}. 
It is important to notice that this way of building higher dimensional lattices does not exhaust all possible lattices.
\begin{table}[h]
\caption{D2Q4 Lattice with $c_{s}=0.60625445810016454$}
\begin{ruledtabular}\begin{tabular}{cl}
$\mathbf{v}_{i}$ & $w_{i} $\\
\hline
(0,0) & $w_{0}^{2}$\\
(0,$\pm$ 1),($\pm$ 1,0) & $w_{0}w_{1}$\\
($\pm$ 1,$\pm$ 1) & $w_{1}w_{1}$\\
(0,$\pm$ 3)($\pm$ 3,0)& $w_{0}w_{3}$\\
($\pm$ 3,$\pm$ 1),($\pm$ 1,$\pm$ 3) & $w_{1}w_{3}$\\
($\pm$ 3,$\pm$3) & $w_{3}^{2}$\\
\label{tableD2Q4}
\end{tabular}\end{ruledtabular}
\end{table}

\clearpage

\section{Algorithmic Details\label{Appendix2}}

\begin{figure}[h!]
\centering{}
\includegraphics[scale=0.27]{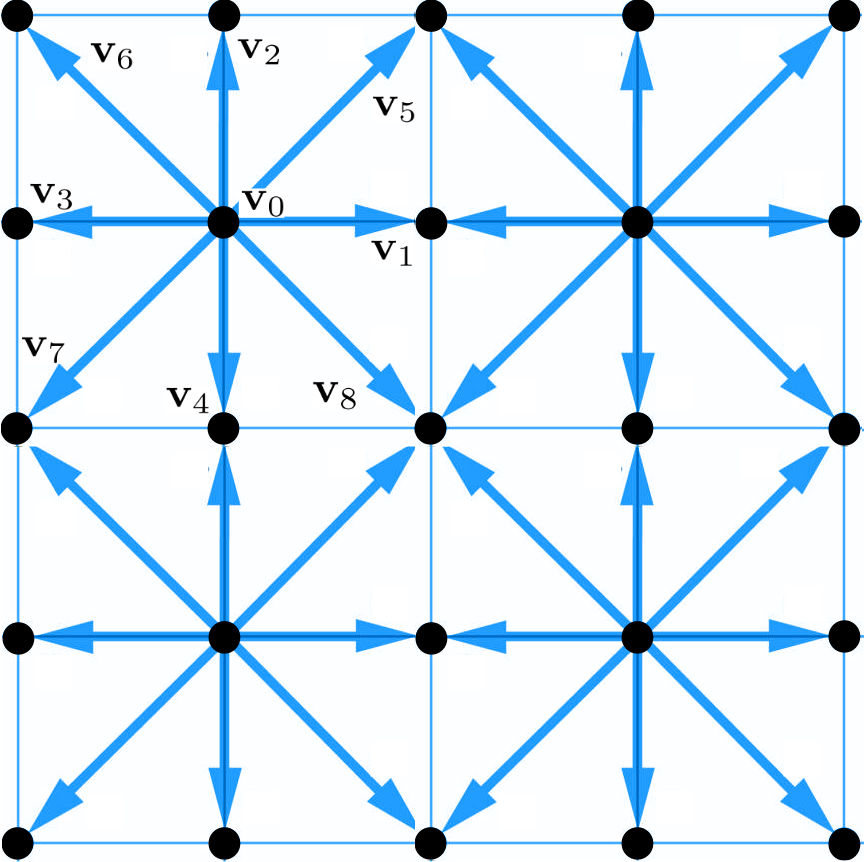}
\caption{(Color online) Scheme of a Lattice Wigner set up. The black circles denote the nodes where the multiple distributions are defined and the arrows show the velocity vectors.}
\label{LatticeGrid}
\end{figure}

The main algorithmic steps of the Lattice Wigner method are similar to those of the Lattice Boltzmann method. 
For reference, consider the lattice shown in Fig.~\ref{LatticeGrid}. At each node (black circle) there are distribution functions $\bar{W}_{i}$, equilibrium distribution function $\bar{W}^{eq}_{i}$ and source term distributions $S_{i}$ where $i=1,2,\dots,N_{q}$. 

Once the initial configuration of the $\bar{W}_{i}$, $\bar{W}^{eq}_{i}$ and $S_{i}$ has been set, the scheme proceeds as follows:

\begin{enumerate}
\item (Collision step) From Eq.~\eqref{lwignerModelEq} calculate for every node and every $i=1,2,\dots,N_{q}$ the so called collision term given by $\bar{W}_{i}^{*}(\mathbf{x},t) =\bar{W}_{i}(\mathbf{x},t)+ \delta t\Omega_{i}+\delta tS_{i} +\frac{\delta t}{2}\left(S_{i}(\mathbf{x},t)-S_{i}(\mathbf{x}-\mathbf{v}_{i}\delta t,t-\delta t)\right)$. 

\item (Streaming step) Observe that Eq.~\eqref{lwignerModelEq} can now be written as $\bar{W}_{i}(\mathbf{x}+\mathbf{v}_{i}\delta t,t+\delta t)=\bar{W}_{i}^{*}(\mathbf{x},t)$. This relation can then be used to update every node for the time step $t+\delta t$.

\item (Macroscopic fields) With the newly updated distribution functions $\bar{W}_{i}(\mathbf{x},t+\delta t)$ the macroscopic fields can be calculated according to Eq.~\eqref{quadratureConstrainEq} and used to update the equilibrium distribution function and source term.
\end{enumerate}

The steps 1,2,3 are iterated until convergence is reached. If the boundary conditions are expressed in terms of distributions, then it is adequate to impose them at the streaming step. If they are expressed in terms of the macroscopic fields then the boundary conditions can be imposed during step 3. For further details see Eq.~\cite{succi2001lattice}

\end{document}